\documentclass[%
 aip,
 amsmath,amssymb,
 reprint,%
]{revtex4-1}

\usepackage{graphicx}
\usepackage{dcolumn}
\usepackage{bm}

\usepackage[utf8]{inputenc}
\usepackage[T1]{fontenc}
\usepackage{mathptmx}
\usepackage{etoolbox}
\usepackage{enumitem}

\makeatletter
\def\@email#1#2{%
 \endgroup
 \patchcmd{\titleblock@produce}
  {\frontmatter@RRAPformat}
  {\frontmatter@RRAPformat{\produce@RRAP{*#1\href{mailto:#2}{#2}}}\frontmatter@RRAPformat}
  {}{}
}%
\makeatother
\begin{document}


\title[Tensor Decomposed Distinguishable Cluster. I.]{Tensor Decomposed Distinguishable Cluster. I. Triples Decomposition}
\author{Charlotte Rickert}
\author{Denis Usvyat}%
\email{rickertc@hu-berlin.de}
\affiliation{ 
Institut für Chemie, Humboldt-Universität zu Berlin, Brook-Taylor-Straße 2, 12489 Berlin, Germany
}%

\author{Daniel Kats}
\email{d.kats@fkf.mpg.de}
\affiliation{%
Max-Planck-Institute for Solid State Research, Heisenbergstraße 1, 70569 Stuttgart, Germany
}%

\date{\today}

\begin{abstract}
We present a cost-reduced approach for the distinguishable cluster
approximation to coupled cluster with singles, doubles and iterative triples
(DC-CCSDT) based on a tensor decomposition of the triples amplitudes. The triples amplitudes and residuals are processed in the 
singular-value-decomposition (SVD) basis. Truncation of the SVD basis
according to the values of the singular values together with the
density fitting (or Cholesky) factorization of the electron repulsion
integrals reduces the scaling of the method to $N^6$, and the DC approximation
removes the most expensive terms of the SVD triples residuals and
at the same time improves the accuracy of the method. 
The SVD basis vectors for the triples are obtained from the approximate CC3 triples density
matrices constructed in an intermediate SVD basis of doubles amplitudes.
This allows us to avoid steps that scale higher than $N^6$ altogether.
Tests against DC-CCSDT and CCSDT(Q) on a benchmark set of chemical reactions
with closed-shell molecules demonstrate that the SVD-error is very small already
with moderate truncation thresholds, especially so when using a CCSD(T) energy
correction. Tests on alkane chains demonstrated that the SVD-error grows
linearly with system size confirming the size
extensivity of SVD-DC-CCSDT within a chosen truncation threshold.  
\end{abstract}

\maketitle



\section{\label{sec:level1} Introduction}

For several decades coupled cluster\cite{bartlett2007coupled, bartlett1991coupled, bartlett2024perspective, crawford2007introduction, zhang2019coupled, bishop1991overview} 
(CC) has been used as an efficient and practical methodological frame to
systematically approach the full configuration interaction (FCI) solution. At
the moment, the most popular flavor of the CC hierarchy is coupled cluster
with singles, doubles and perturbative triples [CCSD(T)]
model.\cite{raghavachari1989fifth} For a very wide range of systems and problems,
 CCSD(T) is
known to deliver chemically accurate (the error below $1\,{\textrm{kcal}}/{\textrm{mol}}$) 
energy differences provided the static correlation is insignificant.\cite{cole1987comparison, stanton1997ccsd} 
An important property of the CC hierarchy of methods that contributes to the
high accuracy and fast and systematic convergence to the exact result is its
size-extensivity. Even though CCSD(T) is computationally very demanding with
an $\mathcal{O}(N^7)$ scaling, applicability 
to big systems of interest can be achieved by the introduction of local approximations 
\cite{schutz2001low, schutz2000local, werner2011,Riplinger2013a,Riplinger2013b,masur2013efficient, schutz2014,Schwilk2015,werner2018,nagy_lccsdt}.

Unfortunately, the perturbative correction (T) fails for systems with even slight multireference
character (for example, already at a homolytic single-bond dissociation).\cite{yang2013ab} Furthermore, it breaks down for metals\cite{shepherd2013many} and is thus expected to be inaccurate for small gap systems.
Iterative triples can solve some of the challenges of CCSD(T) in this
respect,\cite{yang2013ab} which comes, however, with a substantially higher
prefactor and an $\mathcal{O}(N^8)$ scaling.

One of possible ways of including iterative triples contributions is using the
distinguishable cluster (DC) approximated 
CCSDT method (DC-CCSDT).\cite{kats2019distinguishable,
  schraivogel2021accuracy, rishiCan2019} This approximation to full triples neglects some of
the exchange terms of the triples residual equations, while keeping the method (i)
size-extensive, (ii) particle-hole symmetric,\cite{katsParticle2018}
 (iii) exact for three-electron (or three-hole) systems, and (iv) invariant to
 rotations within the occupied and virtual spaces. Furthermore, although DC-CCSDT
 is noticeably less expensive than CCSDT (e.g. in the real-space representation its nominal scaling is $\mathcal{O}(N^7)$), its results are
 usually closer to CCSDT(Q) than those of CCSDT.\cite{kats2019distinguishable,
   schraivogel2021accuracy}
 
Unfortunately, even with the lower scaling, DC-CCSDT is hardly applicable to
large systems without further approximations. As CC in general deals with high
dimensional tensors, one possibility to reduce the computational cost
is to factorize or decompose these tensors. One of the most common examples of
such a decomposition, which is used in various quantum chemistry approaches,
is the density fitting (DF) approximation for the electron repulsion integrals.\cite{baerends1973self, whitten1973coulombic, dunlap1979some, vahtras1993integral, feyereisen1993use, rendell1994coupled} 
Another example is the density matrix
renormalization group (DMRG) approach,\cite{white1992real, schollwock2011density, 
schollwock2005density, chan2008introduction} where the tensor decomposition is
applied to the FCI coefficients. For a review on tensor product approximations
in \textit{ab initio} quantum chemistry we refer to Ref. \onlinecite{szalay2015tensor}.

In this communication, we develop an approximation to DC-CCSDT based on a
tensor decomposition of the triples amplitudes and residuals. 
Initially an idea of reducing the size of the CC amplitudes using a singular value decomposition (SVD) goes back to Bartlett and coworkers who
introduced it for CCD in 2003.\cite{kinoshita2003singular} 
Later, by the same group SVD was applied to the
triples amplitudes in the CCSDT-1 method.\cite{hino2004singular}
In that approach, the triples amplitudes $T^{ijk}_{abc}$ were decomposed as a symmetric
three-index quantity $T_{XYZ}$ contracted with the transformation matrices $U$ referred to as compression coefficients,
\begin{eqnarray}
  T^{ijk}_{abc} &=& \sum_{X Y Z} T_{XYZ} \: U^{iX}_a U^{jY}_b U^{kZ}_c .\label{eq:trip_decomp}
\end{eqnarray}
 This scheme has been recently adopted by
Lesiuk to achieve a reduced
scaling in CCSDT\cite{lesiuk2020implementation,lesiuk2019efficient,lesiuk2021near} and
CCSDT(Q)\cite{lesiukWhen2022} methods. 
In our approach we will also use this type of tensor decomposition for the
triples amplitudes.

We note in passing that also other decomposition schemes have been proposed
within the CC framework. Here one can mention,
for example, a doubles 
amplitudes decomposition in the canonical polyadic tensor format
\cite{benedikt2013tensor} or a 
least-squares tensor hypercontraction (THC) of doubles amplitudes and
integrals.\cite{hohenstein2012communication} These and other schemes or their
combinations have been employed for CCSD, molecular\cite{schutski2017tensor,parrish2019rank,hohenstein2022rank,lesiukQuinticscaling2022} or periodic.\cite{hummel2017low}
In our approach we do not use a tensor decomposition in CCSD itself, but rather 
decompose the relevant block of an approximate two-particle density matrix obtained using converged CCSD doubles amplitudes to reduce scaling in
calculation of the triples two-electron density matrix block, needed to construct
the transformations $U$. With this, the SVD-DC-CCSDT method presented in this work overall scales as $\mathcal{O}(N^6)$.


\section{Theory}


\subsection{Coupled Cluster}

The  CC theory\cite{bartlett2007coupled, bartlett1991coupled,
  bartlett2024perspective, crawford2007introduction, zhang2019coupled,
  bishop1991overview}  features the exponential ansatz for the wave function 
\begin{eqnarray}
      \left| \Psi_{CC} \right> = e^{\hat{T}} \left| \psi_{HF} \right> 
\end{eqnarray}
with the cluster $\hat{T}$ operator defined as
\begin{eqnarray}
  \hat{T} = \hat{T}_1 + \hat{T}_2 + \hat{T}_3 + \ldots,\\
  \hat{T}_n = \left( \frac{1}{n!} \right)^2 T_{AB\ldots}^{IJ\ldots}   \hat{a}_A^\dag \hat{a}_I\hat{a}_B^\dag \hat{a}_J\ldots .\label{eq:clust}
\end{eqnarray}
Here $\psi_{HF}$ is the HF determinant, $T_{AB\ldots}^{IJ\ldots}$ are the
amplitudes, and $\hat{a}^\dag$ and $\hat{a}$ are the creation and annihilation
operators, respectively. We use the standard nomenclature for the orbitals:
 $I,J,K, ...$ denote the occupied HF spin-orbitals and $A,B,C, ...$ the virtual
ones. In (\ref{eq:clust}) and throughout the paper, summation over repeated indices is assumed.
The projected linked coupled cluster equations read:
\begin{eqnarray}
 \left< \psi_{HF} \left| e^{- \hat{T}} \hat{H}_N e^{\hat{T}} \right| \psi_{HF}
  \right> &=& E, \\
 \left< \psi^{I\ldots}_{A\ldots} \left| e^{- \hat{T}} \hat{H}_N e^{\hat{T}} \right| \psi_{HF} \right> &=& 0,\label{eq:ampl}
\end{eqnarray}
where $\psi^{I\ldots}_{A\ldots}$ are the excited determinants, $\hat{H}_N$ -- the
normal ordered Hamiltonian, and $E$ -- the correlation energy. One way to
obtain the residual equations from (\ref{eq:ampl}) is by diagrammatic
techniques,\cite{shavitt2009many} which are also instrumental for specification of the
distinguishable cluster approximation (vide infra).


\subsection{Distinguishable Cluster approximation}
\label{sec:DC-CCSDT_approx}

The DC approximation to CC was initially inspired by the question of how
truncated CC methods can be made more robust in presence of strong electron correlation.
Originally, DC was formulated for the doubles residual equations, giving rise
to a family of methods: DC with singles and doubles
(DCSD), orbital-optimized DCD (ODCD) and Brueckner DCD (BDCD).\cite{kats2013communication, katsCommunication2014} 
Formally, all these approximations omit some of the quadratic exchange terms,
nevertheless keeping exactness for two-electron subsystems, size-extensivity, particle-hole
symmetry, and orbital invariance. As a result, DCSD not only provides a better
qualitative description of strongly correlated systems than CCSD, but is much
more accurate quantitatively quite uniformly across different systems and
properties.\cite{kats2015accurate}

\begin{figure}[h]
  \centering
  \includegraphics[width=6cm]{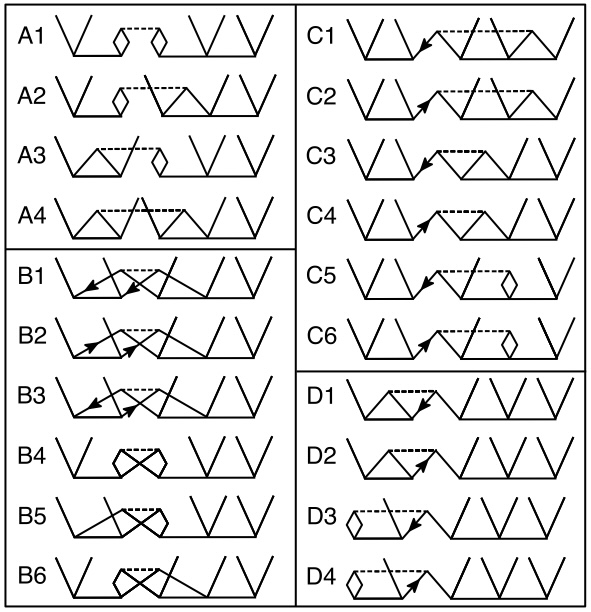}
  \caption{Spin-summed diagrams of $T_2T_3$ contributions to the triples residuals in CCSDT.
  For the DC-CCSDT approximation, the diagrams A1 to A4 are kept, B1 to B6 removed and C1 to C6 as  well as D1 to D4 multiplied with $0.5$.}
\label{fig:T2T3_diagrams}
\end{figure}

Inspired by this success, the DC approximation was extended to CCSDT.\cite{kats2019distinguishable, schraivogel2021accuracy,rishiCan2019} 
To keep the aforementioned constraints and guarantee exactness for
three-electron subsystems, the exchange terms in the doubles residual were
reinstated, while the modification was applied to the CCSDT triples residual
equations.
To illustrate how exactly the DC-CCSDT approximation is defined,\cite{kats2019distinguishable} in Fig.
\ref{fig:T2T3_diagrams} we provide all spin-summed
diagrams of $T_2T_3$ contributions to the triples residuals of CCSDT. The
procedure to obtain the spin-summed residual equations is described below in
section \ref{sec:resid}. Here we just point out how one can obtain the DC-CCSDT
approximation from CCSDT diagrammatically: In DC-CCSDT the terms corresponding
to the diagrams A1 to A4 remain unchanged, the terms B1 to B6 are fully
omitted, and the terms corresponding to diagrams C1 to C6 and D1 to D4 are
additionally multiplied by a factor $1/2$.

\subsection{Tensor decompositions}


\subsubsection{Factorization of electron repulsion integrals}

The electron-repulsion integrals (ERIs) are factorized as symmetric contractions of 3
index quantities. In a standard workflow employing explicit basis sets, this is
achieved by the density fitting approximation:\cite{baerends1973self, whitten1973coulombic, dunlap1979some, vahtras1993integral, feyereisen1993use, rendell1994coupled} 
\begin{eqnarray}
    v_{pq}^{rs} \approx v_p^{rP} \left[v^{-1}\right]_{PQ} v_q^{sQ},
\end{eqnarray}
where $v_{pq}^{rs}$, $v_p^{qP}$ and $v^{PQ}$ are 4-index, 3-index and
2-index ERIs, respectively:
\begin{eqnarray}
v_{pq}^{rs}&=&\int d\textbf{r}_1 d\textbf{r}_2 \frac{\phi^*_p(\textbf{r}_1) \phi^r(\textbf{r}_1) \phi^*_q(\textbf{r}_2)\phi^s(\textbf{r}_2)}{|\textbf{r}_1-\textbf{r}_2|} ,\\  
v_p^{qP}&=& \int d\textbf{r}_1
  d\textbf{r}_2\frac{\phi^*_p(\textbf{r}_1) \phi^q(\textbf{r}_1)
  \phi^P(\textbf{r}_2)}{|\textbf{r}_1-\textbf{r}_2|},\\
  v^{PQ}&=& \int d\textbf{r}_1
  d\textbf{r}_2\frac{\phi^P(\textbf{r}_1) \phi^Q(\textbf{r}_2)}{|\textbf{r}_1-\textbf{r}_2|} .
\end{eqnarray}
Here the auxiliary orbitals are denoted by capital letter indices $P$, $Q$,
while $p$, $q$, $r$ and $s$ denote general (occupied or virtual) spatial orbitals.
A symmetric decomposition 
\begin{eqnarray}
  v_{pq}^{rs} \approx v_{p}^{rL} v_{q}^{sL}\label{eq:ERIs_fact}
\end{eqnarray}
is straightforwardly obtained by 
\begin{eqnarray}
  v_{p}^{qL} = v_p^{qP}\left[v^{-{1 \over 2}}\right]_{P}^{L},
\end{eqnarray}
where $L$ can correspond to the auxiliary basis functions or to eigenvectors of $v^{PQ}$, etc.
Alternatively, the 4-index ERI $v_{pq}^{rs}$ can be obtained in a separate
calculation and passed to the \texttt{ElemCo.jl} code via an FCIDUMP-interface (together
with the one-electron Hamiltonian).\cite{knowlesDeterminant1989}
This opens a possibility to apply a high-level correlated treatment to an
important orbital subspace only, restricted in the energy or the direct-space
representation. Such an approach becomes especially appealing in studying local
effects in large molecules\cite{Mata:08} or periodic
systems.\cite{mullan21,christlmaier21,Rob24} With the 4-index ERIs of the FCIDUMP-interface, density fitting is
not applicable. In this case, the factorization in Eq.~(\ref{eq:ERIs_fact}) is performed
directly by a diagonalization of the ERI matrix and removal of the eigenvectors
corresponding to eigenvalues smaller than a given threshold.


\subsubsection{Triples amplitudes decomposition}\label{sec:decom_Trip}
As is mentioned in the introduction, we employ the decomposition
(\ref{eq:trip_decomp}) for the triples amplitudes. With the transformation
matrices $U$ this decomposition can be also seen as a formal transformation of the
triples amplitudes from the SVD basis $T_{XYZ}$ to the orbital basis
$T^{ijk}_{abc}$. The transformation connects {\it one} index in the SVD space and {\it two} indices -- one occupied and one virtual -- in the orbital space. 

In the SVD-DC-CCSDT method the triples amplitudes are varied
directly in the SVD basis, such that the residual $R_{XYZ}$ -- again in the SVD basis --  goes to zero. 
Importantly, each $U$-transformation can be applied to one SVD index independently of the others.  
This provides a flexibility, such that the transformation from SVD to the orbital basis in each diagram can
be restricted only to the indices that are contracted with the Hamiltonian, while the others can stay in the SVD basis through the complete contraction. 

\begin{figure}[h]
  \centering

        \includegraphics[width=6.5cm]{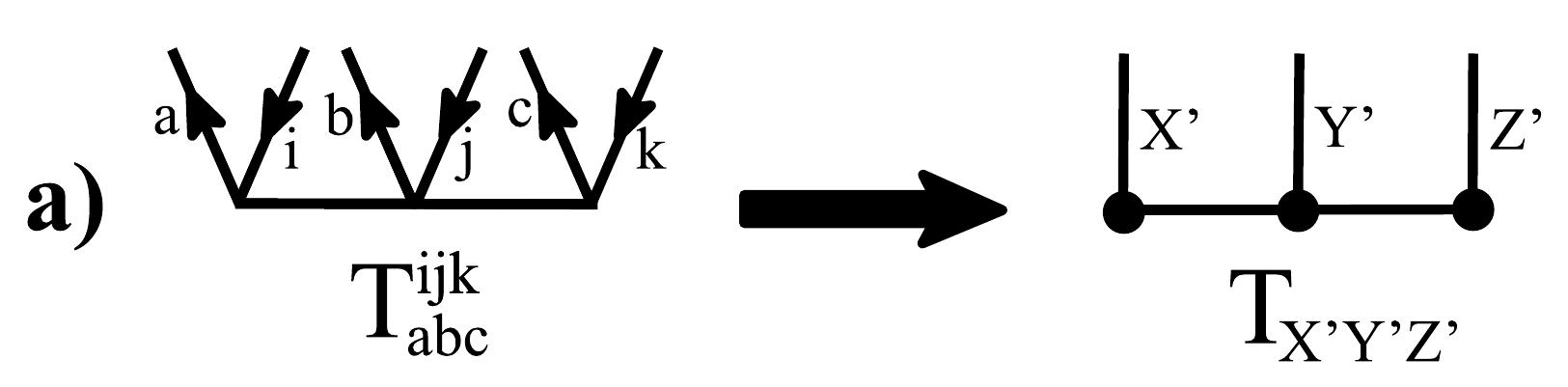} 
        \centering
        \includegraphics[width=1.5cm]{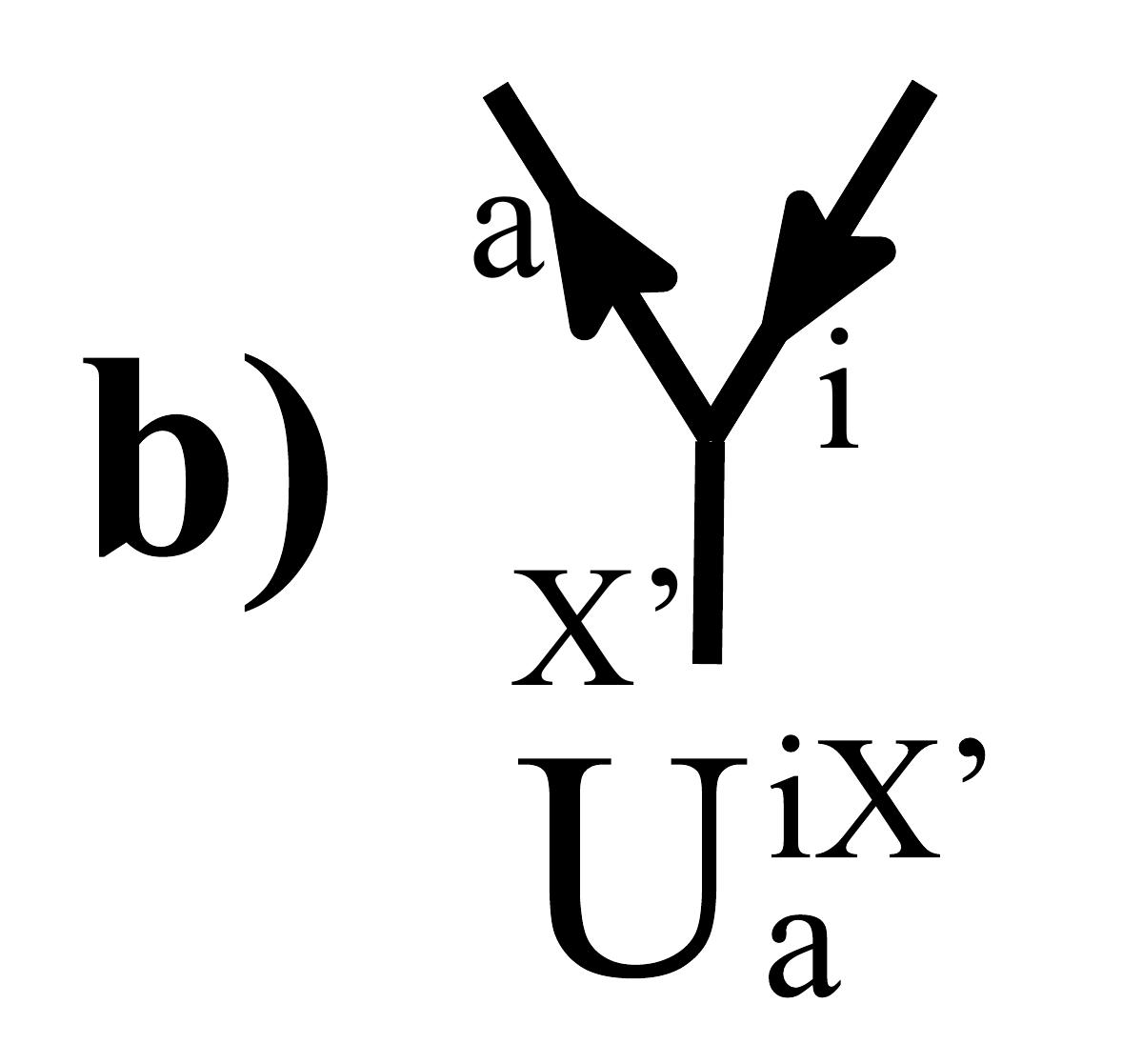} \\
    

        \includegraphics[width=6.5cm]{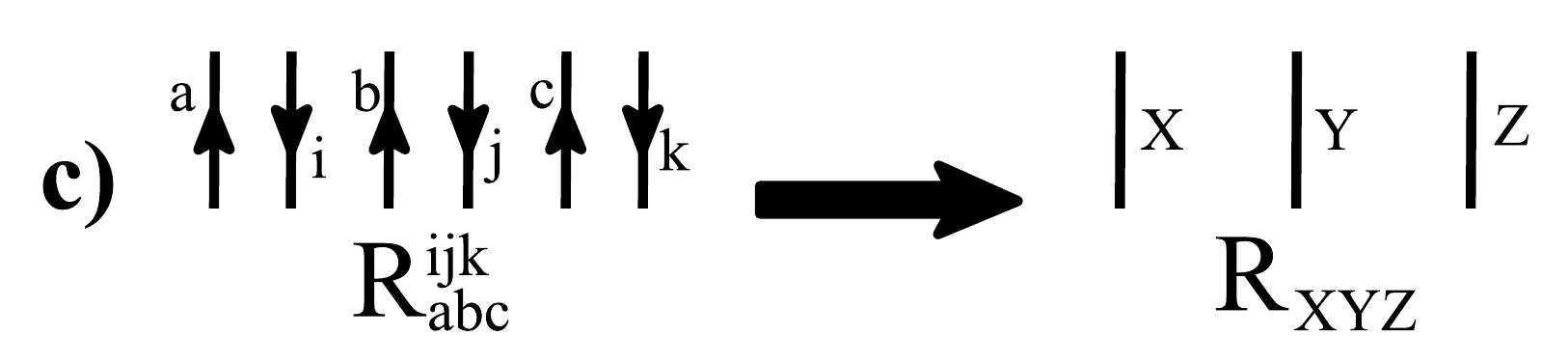} 
        \includegraphics[width=1.5cm]{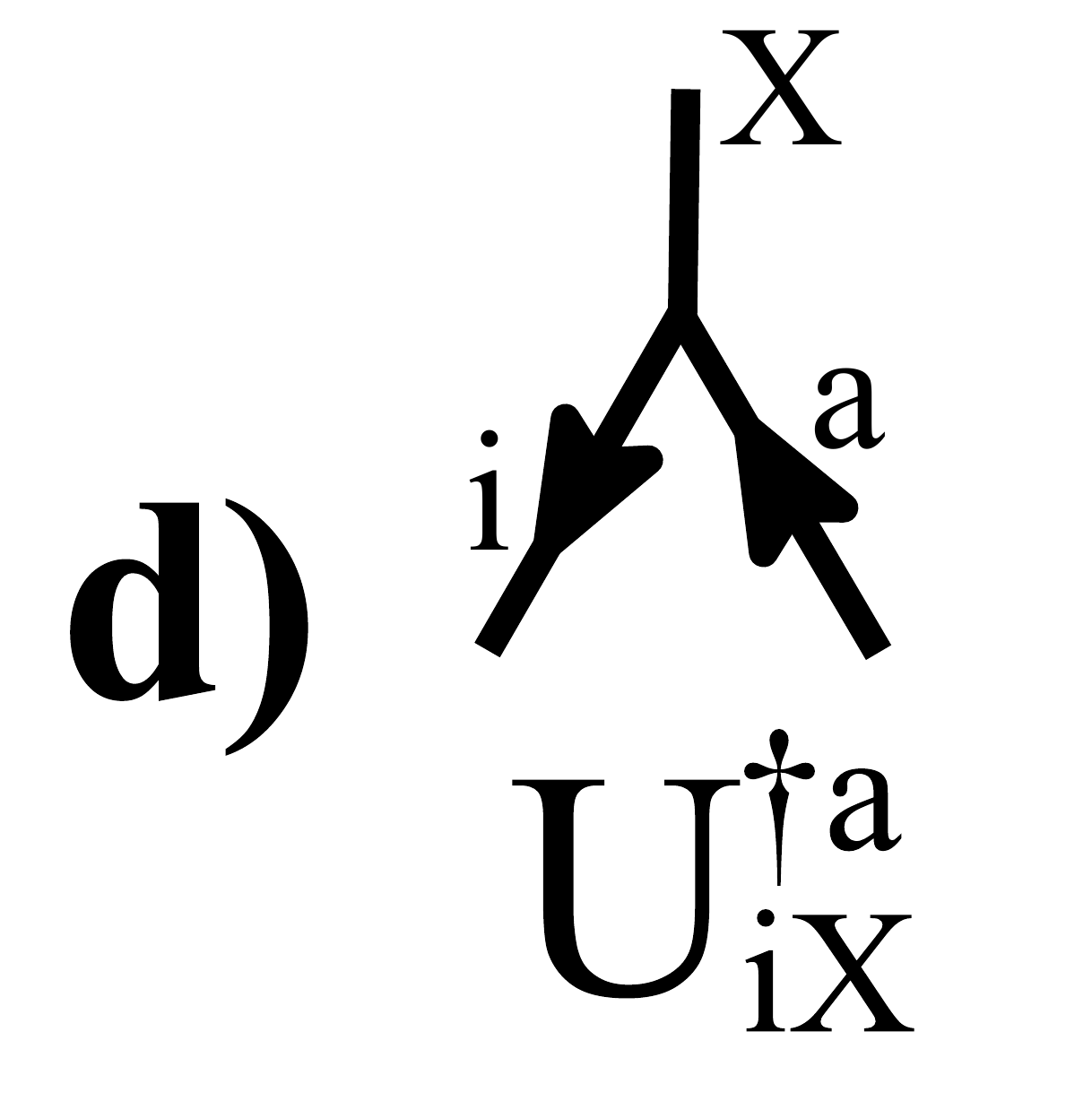} 
  \caption{Diagrammatic elements for the SVD-CCSDT formalism: the triples
    cluster operator in the SVD basis (a), the transformation from the SVD to
    the orbital basis (b), the residual in SVD basis (c), the transformation
    from the orbital to the SVD basis (d).}\label{fig:Diagr_elem}
\end{figure}

In order to formalize the reduction of the index number and the transformation between the spaces in the SVD-CC theory we introduce new elements in the diagrammatic representation. 
Firstly it concerns the triples amplitudes $T_{XYZ}$, for which each pair of arrowed particle and hole lines is substituted with an unarrowed SVD line, see Fig. \ref{fig:Diagr_elem}a. 
A transformation of an SVD index in a pair of orbital indices via a $U$ matrix will be denoted by a vertex, an SVD-line below it and a particle and a hole line above it, see Fig. \ref{fig:Diagr_elem}b.

The same principles are applied to the triples residuals that at the end also appear in the SVD basis, which in the diagrammatic representation correspond to the three unarrowed lines, see Fig. \ref{fig:Diagr_elem}c. 
The residuals transform conversely with respect to amplitudes (as covariant and contravariant vectors).\cite{katsUse2008} 
Therefore the transformation $U$ that brings the amplitudes from SVD to the orbital basis, 
if subjected to transposition and complex conjugation $U^\dagger$, transforms residuals conversely from the orbital basis to SVD.
The diagrammatic representation of $U^\dagger$ is given in Fig. \ref{fig:Diagr_elem}d.
$U$ and $U^\dagger$ are orthogonal to each other, $U^{\dagger a}_{iX} U^{iY}_a = \delta_X^Y$.

\begin{figure}[h]
  \centering
  \includegraphics[width=9cm]{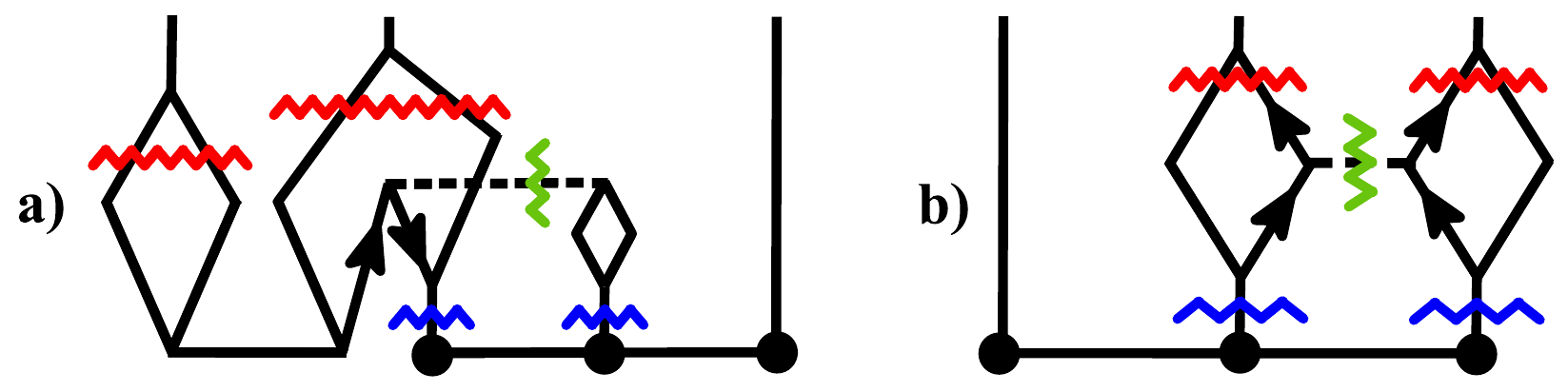}
  \caption{Diagram C6 (a) and a triples-ladder diagram (b) contributing to the
    triples residuals are shown in the SVD format.
  The blue wavy lines indicate the insertion of the transformation $U$ (Fig. \ref{fig:Diagr_elem}b) from the SVD to the orbital
  basis, the red wavy lines -- of the back transformation $U^\dagger$
  (Fig. \ref{fig:Diagr_elem}d) to the SVD
  basis. The green wavy lines show the factorization of the ERIs, Eq.~(\ref{eq:ERIs_fact}).}\label{fig:diagrams_svd}
\end{figure}

With these rules, we can diagrammatically represent the matrix elements within the SVD-DC-CCSDT formalism. 
For example Fig. \ref{fig:diagrams_svd} shows two exemplary diagrams from the triples residual:  
the diagram C6 from Fig. \ref{fig:T2T3_diagrams} and a triples ladder diagram. 
The triples amplitudes are given in the SVD basis. Then, as marked by the horizontal blue cuts, for the relevant lines the $U$ transformations to the orbital basis are introduced. 
After contraction with the Hamiltonian the orbital lines of the residual are transformed back to the SVD basis, as marked by the horizontal red cuts. Finally, the vertical green 
cuts denote the decomposition of the ERIs, Eq.~(\ref{eq:ERIs_fact}).

\subsection{Residual equations}\label{sec:resid}

To achieve an efficient implementation, we obtain the SVD-DC-CCSDT equations in the spin-summed formalism.
Given the singlet excitation operator
\begin{equation}
  \hat{E}^a_{i} = \hat{a}^\dagger_{a \alpha} \hat{a}_{i \alpha} + \hat{a}^\dagger_{a \beta} \hat{a}_{i \beta}
\end{equation}
where the indices $i$ ($j$, $k$, ...) and  $a$ ($b$, $c$, ...) denote the occupied and virtual spatial orbitals, respectively, one defines the standard singly, doubly and triply excited configuration state functions (CSFs):
\begin{eqnarray}
    \left| \Phi_i^a \right> &=& \hat{E}^a_{i} \left| \psi_0 \right> = \left| \psi^a_i \right> + \left| \psi^{\bar{a}}_{\bar{i}} \right>\\
  \left| \Phi_{ij}^{ab} \right> &=&{1 \over 2} \hat{E}^a_{i} \hat{E}^b_{j} \left| \psi_0 \right> 
  \nonumber\\
 & =& {1 \over 2}\left[ \left| \psi^{ab}_{ij} \right> 
  + \left| \psi^{\bar{a} \bar{b}}_{\bar{i} \bar{j}} \right>
  + \left| \psi^{a \bar{b}}_{i \bar{j}} \right>
  + \left| \psi^{\bar{a} b}_{\bar{i} j} \right>\right],\\
    \left| \Phi_{ijk}^{abc} \right> &=&  {1 \over 6}\hat{E}^a_{i} \hat{E}^b_{j} \hat{E}^c_{k} \left| \psi_0 \right> \nonumber
    \\
   & =& {1 \over 6}\left[\left| \psi^{abc}_{ijk} \right> 
    + \left| \psi^{\bar{a} \bar{b} \bar{c}}_{\bar{i} \bar{j} \bar{k}} \right>
    + \left| \psi^{a b \bar{c}}_{i j \bar{k}} \right>
    + \left| \psi^{\bar{a} \bar{b} c}_{\bar{i} \bar{j} k} \right>\right.\nonumber
    \\
   & &+ \left.\left| \psi^{a \bar{b} c}_{i \bar{j} k} \right>
    + \left| \psi^{\bar{a} b \bar{c}}_{\bar{i} j \bar{k}} \right>
    + \left| \psi^{\bar{a} b c}_{\bar{i} j k} \right>
    + \left| \psi^{a \bar{b} \bar{c}}_{i \bar{j} \bar{k}} \right>\right]\label{eq:trip_CSF}
\end{eqnarray}
and the corresponding amplitudes
\begin{eqnarray}
  \hat{T}_1 &=& T^i_a \hat{E}^a_{i},\\
  \hat{T}_2 &=& {1\over 2} T^{ij}_{ab} \hat{E}^a_{i} \hat{E}^b_{j},\\
  \hat{T}_3 &=& {1\over 6}T^{ijk}_{abc} \hat{E}^a_{i} \hat{E}^b_{j} \hat{E}^c_{k}.
\end{eqnarray}
The bars on top of the orbital indices indicate the $\beta$-spin.

The singles contributions to the residuals are implemented using the similarity transformation on the Hamiltonian \cite{koch1996integral}
\begin{eqnarray}
\tilde{H}_N = e^{-\hat{T}_1} \hat H_N e^{\hat{T}_1},
\end{eqnarray}
which implies ``dressing'' of the Fock matrices and the ERIs factorization tensors $v_{p}^{qL}$ with the singles amplitudes. 
The dressed quantities are denoted with a hat.

\subsubsection{Singles and doubles residual equations}

For the bra-projection in the singles and doubles residual equations we use the contravariant CSFs:\cite{pulay1984efficient,knowles1993coupled} 
\begin{eqnarray}
\widetilde{\Phi}_i^a &=& \frac{1}{2}\left(\psi_{i}^{a} + \psi_{\bar{i}}^{\bar{a}}\right),\\
  \widetilde{\Phi}_{ij}^{ab} &=& \frac{1}{3}\left(\psi_{ij}^{ab} 
  + \psi_{\bar{i}\bar{j}}^{\bar{a}\bar{b}}
  + 2 \psi_{i\bar{j}}^{a\bar{b}}
  + 2 \psi_{\bar{i}j}^{\bar{a}b}
  + \psi_{i\bar{j}}^{b\bar{a}}
  + \psi_{\bar{i}j}^{\bar{b}a} \right).
\end{eqnarray}

The pure singles and doubles terms with such a bra-projection can be found in the \texttt{ElemCo.jl} documentation.\cite{elemcojl-docs} 
Here we focus only on the terms that involve the triples. To get the actual expressions for these terms within the SVD-DC-CCSDT we adopt the following protocol:
\begin{enumerate}
\item Calculate the algebraic expressions for the matrix element at question
  with the program \texttt{Quantwo},\cite{quantwo} which
  symbolically evaluates all the terms for the given bra- and ket-CSFs
  using the Wick's theorem.
\item Replace the triples amplitudes with the SVD-decomposed form, Eq.~(\ref{eq:trip_decomp}),
and the ERIs with the factorized form, Eq.~(\ref{eq:ERIs_fact}).
\item Factorize the equations. For this the diagrammatic representation can be used,
as discussed in section \ref{sec:decom_Trip} (see Figs \ref{fig:Diagr_elem} and \ref{fig:diagrams_svd}).
\end{enumerate}
After introducing intermediates, this yields for triples contributions to the singles residual:
\begin{eqnarray}
R_a^i &\leftarrow& U_a^{iX} \left[T_{XYZ} \left(2A^{YL} A^{ZL} - B^{YZ}\right)\right]\nonumber
\\
&&- U_a^{jY} \Big[2 {B}_{j}^{iXL} V_{XY}^{L}- G_{jY}^{i}\Big],\label{eq:sing_int1}  
\end{eqnarray}
with intermediates
\begin{align}
&A^{XL}={B}_{i}^{iXL}, \hspace{1cm} &{B}_{i}^{jXL} = v_i^{aL} U_a^{jX},\nonumber\\
&B^{XY}={B}_{i}^{jXL} {B}_{j}^{iYL}, \hspace{1cm} 
&w_k^{dX} = v_l^{dL} {B}_k^{lXL},\nonumber\\
&V_{XY}^{L} = T_{XYZ} A^{ZL}, \hspace{1cm} &G_{jY}^{i} = T_{dYZ}^{i} w_j^{dZ},\nonumber\\
&T_{aYZ}^{i} = T_{XYZ} U_a^{iX}, .&\label{eq:sing_int2}
\end{align}
The parentheses denote the contraction order.
The triples contributions to the doubles residual read:
\begin{eqnarray}
\label{eq:doub_int1}
R_{ab}^{ij} &\leftarrow & \mathcal{S} \left(ij,ab\right) \Big\{ 
- \left(\bar B_a^{iLZ} - B_a^{iLZ}\right) V_{bZ}^{jL}
\\
&&+U_b^{jZ} \Big[2\left(\bar B_a^{iLY} - B_a^{iLY}\right) V_{YZ}^{L}
+ T_{aYZ}^{i} \left(\hat{f}_k^c U_c^{kY}\right)
\nonumber\\
&&  - T_{XYZ} \left(U_c^{iY}\left(\bar B_a^{kLX} v_k^{cL}\right) 
- U_a^{kY} \left(B_c^{iLX} v_k^{cL} - \hat{f}_k^c U_c^{iX}\right)\right)\Big]
 \Big\},
\nonumber
\end{eqnarray}
with intermediates
\begin{align}
&B_a^{iLX} = \hat{v}_j^{iL} U_a^{jX},  \hspace{1cm} &\bar B_a^{iLX} = \hat{v}_a^{bL} U_b^{iX},\nonumber\\
&V_{bZ}^{jL} = {B}_k^{jYL} T_{bYZ}^{k}, \label{eq:doub_int2}
\end{align}
and the symmetrization operator
\begin{equation}
\mathcal{S} \left(ij,ab\right) X_{ab}^{ij} = X_{ab}^{ij} + X_{ba}^{ji}.
\end{equation}

The correlation energy $E$ is calculated as usually using the singles and doubles 
amplitudes:
\begin{eqnarray}
E&=& \left(2 T_a^i T_b^j - T_b^i T_a^j\right) v_{ij}^{ab} + 2 T_a^i f^a_i
\nonumber\\
&&+ (2T_{ab}^{ij} - T_{ba}^{ij}) v_{ij}^{ab}.
\end{eqnarray}

\subsubsection{Triples residual equations}


For the triply excited covariant CSF the permutation of virtual orbital indices 
leads to a linear dependency for one CSF on the other five CSFs.\cite{schutz2000low,paldus1988clifford} 
As a result a fully orthogonal contravariant CSF for the bra-projection cannot be constructed for triples. 
We employ the contravariant triples CSFs of Ref. \onlinecite{schutz2000low} (apart from the 1/6 prefactor, 
which in our case is absorbed in the covariant CSF $\Phi_{ijk}^{abc}$, eq. (\ref{eq:trip_CSF})). 
To ease the evaluation of the matrix elements with the \texttt{Quantwo} program\cite{quantwo} we express it via the covariant ones as
\begin{eqnarray}
  \widetilde{\Phi}_{abc}^{ijk}={\Phi}_{abc}^{ijk}-{\Phi}_{cab}^{ijk}.
\end{eqnarray}
Importantly, due to the overlap properties of the chosen contravariant and covariant CSFs (see eq. (8) of Ref. \onlinecite{schutz2000low})
\begin{eqnarray}
  \left<\widetilde{\Phi}_{abc}^{ijk}|{\Phi}^{abc}_{ijk}\right>&=&1\nonumber\\
  \left<\widetilde{\Phi}_{abc}^{ijk}|{\Phi}^{acb}_{ijk}\right>&=&\left<\widetilde{\Phi}_{abc}^{ijk}|{\Phi}^{bac}_{ijk}\right>=\left<\widetilde{\Phi}_{abc}^{ijk}|{\Phi}^{bca}_{ijk}\right>=\left<\widetilde{\Phi}_{abc}^{ijk}|{\Phi}^{cba}_{ijk}\right>\nonumber\\&=&0\nonumber\\
  \left<\widetilde{\Phi}_{abc}^{ijk}|{\Phi}^{cab}_{ijk}\right>&=&-1\label{eq:overlaps}
\end{eqnarray}
the matrix elements with such a contravariant CSF bra-projection will consist
of two sets of terms. The second set can be obtained from the first one by the
index-permutation $abc\rightarrow cab$ and multiplication with -1. This can be expressed by the antisymmetrized permutation operator
\begin{eqnarray}
  \mathcal{P}(abc;cab)F(a,b,c)&=&F(a,b,c)-F(c,a,b)
\end{eqnarray}  
For an illustration, consider a simple matrix element $\left<\widetilde{\Phi}_{abc}^{ijk} | \hat{T} _3| \psi_0 \right>$. Due to the overlap relations (\ref{eq:overlaps})
\begin{equation}
  \left<\widetilde{\Phi}_{abc}^{ijk} | \hat{T} _3| \psi_0 \right>= T^{ijk}_{abc} - T^{ijk}_{cab} = \mathcal{P}(abc;cab) T^{ijk}_{abc}.
\end{equation}
For this effect in the (T) residual we refer to eq. (17) in Ref. \onlinecite{schutz2000low}. 
Similarly, the full CCSDT triples residual $\bar{R}^{ijk}_{abc}$ can be represented by two terms $R^{ijk}_{abc}$ and $R^{ijk}_{cab}$, 
one obtained by an index permutation of the other:
\begin{eqnarray}
&&  \left<\tilde{\Phi}_{abc}^{ijk} |
  \exp\left(-\hat{T}_2-\hat{T}_3\right)\tilde{H}_N\exp\left(\hat{T}_2+\hat{T}_3\right)|
   \psi_0 \right>=\bar{R}^{ijk}_{abc}\nonumber\\
  &&\,\,\,\,\,\,\,\,\,\,\,\,\,\,\,\,\,\,=R^{ijk}_{abc} - R^{ijk}_{cab} = \mathcal{P}(abc;cab) R^{ijk}_{abc}
\end{eqnarray}
Importantly, when the $R^{ijk}_{abc}$ residual vector is zero the permuted term $R^{ijk}_{cab}$ is zero too. 
Therefore, it is sufficient to solve the equations:
\begin{equation}
 R^{ijk}_{abc}=0
\end{equation}
with just half of the terms of the initial residual  $\bar{R}^{ijk}_{abc}$.

To define the triples residual equations to be solved we introduce a formal
operator $\mathcal{P}^{-1}(abc;cab)$ as an inverse of the initial operator
\begin{equation}
\mathcal{P}^{-1}(abc;cab)\mathcal{P}(abc;cab)=1.
\end{equation}
It has an obvious property:
\begin{equation}
\mathcal{P}^{-1}(abc;cab)\left[F(a,b,c)-F(c,a,b)\right]=F(a,b,c).\label{eq:perm_inv}
\end{equation}
With this operator we can define our bra-projection for the triples residual equations as $\mathcal{P}^{-1}(abc;cab)\left<\tilde{\Phi}_{abc}^{ijk}\right|$,
where the operator $\mathcal{P}^{-1}(abc;cab)$ is meant to be applied not to the bra-function only but rather to the complete matrix element. With that we obtain 
\begin{eqnarray}
  &&\mathcal{P}^{-1}(abc;cab)\left<\tilde{\Phi}_{abc}^{ijk} \right|\nonumber\\
  &&\,\,\,\,\,\,\,\,\,\,\,\,\,\,\,\,\,\left. \exp\left(-\hat{T}_2-\hat{T}_3\right)\tilde{H}_N\exp\left(\hat{T}_2+\hat{T}_3\right)| \psi_0 \right>=R^{ijk}_{abc} .
\end{eqnarray}
Technically, application of the operator $\mathcal{P}^{-1}(abc;cab)$ to a set of terms, obtained by expanding the matrix element, involves 
a polynomial division of these terms by $\mathcal{P}(abc;cab)$, represented as a linear combination of virtual index permutations: $(123)-(312)$. 

Below we summarize the workflow for the determination of the terms in the triples residual equations of SVD-DC-CCSDT.
\begin{enumerate}
\item Calculate the algebraic expressions for the matrix element $\left<\tilde{\Phi}_{abc}^{ijk} | \exp\left(-\hat{T}_2-\hat{T}_3\right)\tilde{H}_N\exp\left(\hat{T}_2+\hat{T}_3\right)| \psi_0 \right>$ using the program \texttt{Quantwo}.\cite{quantwo}
\item Apply the operator $\mathcal{P}^{-1}(abc;cab)$ to the obtained terms using \texttt{Quantwo}. 
 \item Identify the  B-, C- and D-diagrams of Fig. \ref{fig:T2T3_diagrams} and scale or remove them to define the DC-CCSDT approximation.
 It can also be done in \texttt{Quantwo} by subtracting the corresponding terms.
 \item Replace the triples amplitudes with the SVD-decomposed form, Eq.~(\ref{eq:trip_decomp}),
 the ERIs with the factorized form, Eq.~(\ref{eq:ERIs_fact}), and multiply the residual equation with $U^{\dagger a}_{iX} U^{\dagger b}_{jY} U^{\dagger c}_{kZ}$.
 \item Factorize the terms considering the orthogonality of $U$ using common intermediates.  
 The latter can be identified by searching for identical diagrammatic fragments after slicing the diagrams according to Fig. \ref{fig:diagrams_svd}.   
\end{enumerate}

The triples residual equations in the SVD basis, obtained using this protocol, read: 
\begin{eqnarray}
  R_{XYZ}& =& Q_{XYZ} + Q_{YXZ} + Q_{XZY} + Q_{ZYX} + Q_{ZXY} + Q_{YZX} \nonumber\\
   &&+ q_{XYZ} + q_{YXZ} + q_{ZYX}\label{eq:resid_trip}
\end{eqnarray}
with the main intermediates
\begin{eqnarray}
  Q_{XYZ} &=& U^{\dagger b}_{jY} \Big[T_{bX}^l X_{lZ}^j - X_{bZ}^d T_{dX}^j \nonumber\\
  &&+T_{bY'Z}^{l} \left(U_d^{jY'}  \left(v_l^{dL} Y_X^L \right)\right)\Big]  \label{eq:big_int1}
\end{eqnarray}
and
\begin{eqnarray}
  q_{XYZ} &=& T_{X'YZ} \Big[U^{\dagger a}_{iX} \Big( \left(\hat{f}_l^i + 0.5 v_l^{dL} V_d^{iL}\right)U_a^{lX'}
  \nonumber\\
 & &- \left(\hat{f}_a^d - 0.5 v_l^{dL} V_{a}^{lL}\right) U_d^{iX'}
  \nonumber\\
 & &+ \left(\hat{v}_l^{iL} \hat{v}_a^{dL}\right) U_d^{lX'}\Big)
 -2 Y_X^L A^{X'L}\Big]
  \nonumber\\
 & &- T_{XY'Z'} \left( W_Y^{Y'L} W_Z^{Z'L}\right) .
\label{eq:big_int2}      
\end{eqnarray}
The further intermediates used in these expressions are 
\begin{eqnarray}
X_{bZ}^d &=& \hat{v}_b^{dL} Y_Z^L - \hat{f}_l^d T_{bZ}^l 
 - U^{lX}_{b} \left(v_l^{dL} \tilde V_{ZX}^{L}\right)
\nonumber\\
&& 
+ \left(\hat v_{lk}^{di} T^{lk}_{ba} - \hat v_{la}^{dc} T^{li}_{bc}\right) U^{\dagger a}_{iZ}
- \hat v_{lb}^{dc} T_{cZ}^l 
\nonumber\\
&& + 0.5 T_{bYZ}^{k} w_k^{dY},\nonumber\\
X_{lZ}^j&=& \hat{v}_l^{jL} Y_Z^L
 + U^{jX}_{d} \left(v_l^{dL} \tilde V_{ZX}^{L}\right)
 \nonumber \\
 && + \left(\hat v_{la}^{dc} T^{ji}_{dc} - \hat v_{lk}^{di} T^{jk}_{da}\right) U^{\dagger a}_{iZ}
  - \hat v_{lk}^{dj} T^k_{dZ}
 \nonumber \\
  && - 0.5 G_{lZ}^{j} \label{eq:int2}
\end{eqnarray}
and
\begin{equation}
\begin{split}
  &V_a^{iL} = v_k^{cL}  \widetilde{T}_{ac}^{ik} \hspace{1.5cm} 
  Y_X^L = U^{\dagger a}_{iX} \left(\hat{v}_a^{iL} + V_a^{iL} \right)
  \\
&T_{aX}^i = U^{\dagger b}_{jX}  T_{ab}^{ij} \hspace{1.3cm} 
\widetilde{T}_{ab}^{ij} = 2  T_{ab}^{ij} - T_{ba}^{ij}
\\
&\hat v_{la}^{dc} = v_l^{dL} \hat v_a^{cL} \hspace{1.5cm}
\hat v_{lk}^{di} = v_l^{dL} \hat v_k^{iL}
\\
&W_{Y}^{Y'L} =  U^{\dagger b}_{jY}  \left(\bar B_b^{jLY'} - B_b^{jLY'}\right)\\
&\tilde V_{ZX}^{L} = V_{ZX}^{L} - 0.5 V_{cX}^{kL} U^{\dagger c}_{kZ}
  .\label{eq:int4}
\end{split}
\end{equation}

None of the contractions in eqs. (\ref{eq:sing_int1})-(\ref{eq:doub_int2}) and (\ref{eq:big_int1})-(\ref{eq:int4}) involve more than six different indices. 
This means that the tensor decompositions reduce the scaling of the DC-CCSDT method, 
which conventionally is ${\cal O}(N^8)$, to ${\cal O}(N^6)$, provided the number of SVD basis functions 
grows linearly with the system size (vide infra). 
Furthermore, some particularly nasty terms become inexpensive, 
like e.g. the 4-external ladder diagram of the triples amplitudes 
(the last term in eq. (\ref{eq:big_int2})), which is reduced to even  $N^5$. 

The most expensive terms in the SVD-DC-CCSDT residual equations scale as $N_o^2 N_v N_X^2 N_L$,
where $N_o$ and $N_v$ are the numbers of occupied and virtual orbitals,
and $N_X$ and $N_L$ are the numbers of SVD and density-fitting basis functions, respectively.
Assuming $N_v \approx N_X$ (vide infra), the most expensive terms in SVD-DC-CCSDT are 
only a factor of $N_L/N_v$ more expensive than the most expensive term in CCSD 
(the 4-external ladder diagram with the $N_o^2 N_v^4$ scaling).
This is a significant improvement over the SVD-CCSDT method (especially for large basis sets),
where the most expensive term scales as $N_o N_v^2 N_X^3$, see Ref. \onlinecite{lesiuk2020implementation}.

An additional approximation can be introduced by calculating the $V_{aX}^{iL}$ intermediate 
in an SVD basis, which would reduce the scaling of the most expensive terms to $N_o^2 N_v N_X^3$,
\begin{align}
  \label{eq:proj}
  &V_{aX}^{iL} \approx \bar V_{X\tilde Z}^{L} \tilde U^{i\tilde Z}_{a} ,\\
  &\bar V_{X\tilde Z}^{L} = \left(\left(\tilde U^{\dagger b}_{j\tilde Z} U^{kY}_{b} \right) T^{j}_{cXY}\right) v_k^{cL}.
\end{align}
In principle, one can use the normal SVD space for $\tilde Z$, but since the $\tilde Z$ SVD space has to  
cover the triples SVD space (for $\tilde V_{XZ}^{L}$ intermediate, Eq.~(\ref{eq:int4})),
and the doubles space (for $\tilde U^{i\tilde Z}_a$ projection in the doubles residual equations, Eq.~(\ref{eq:doub_int1})),
we use a combined SVD space constructed by expanding the triples SVD space by important orthogonal contributions 
from the doubles SVD space (more precisely, from the SVD space from contravariant doubles amplitudes, cf. \ref{sec:svd_basis}).

\subsection{Construction of the SVD basis}
\label{sec:svd_basis}

In order to achieve the ${\cal O}(N^6)$ scaling of the SVD-DC-CCSDT method at every stage of the calculation, the construction of the singular vectors $U^{iX}_{a}$ should scale not higher than that. 
Generally the singular values and singular vectors can be obtained by diagonalizing an approximate triples' two-particle density matrix block $D^{aj}_{ib}$. 
This diagonalization scales as $N^6$, but the computation of such density matrix scales as $N^8$ even at the (T)-level. 
To reduce this scaling we use a two-step procedure. 

First, we obtain the relevant block of the approximate two-particle density matrix from converged CCSD doubles amplitudes

\begin{equation}
\mathcal{D}^{aj}_{ib} = T^{\dagger ac}_{ik} T^{jk}_{bc},
\end{equation}
and eigendecompose it,
\begin{equation}
\mathcal{D}^{aj}_{ib} \approx \bar{U}_{i\tilde{\bar X}}^{\dagger a} \bar{D}_{\tilde{\bar X}}
\bar{U}^{j\tilde{\bar X}}_{b}.
\label{eq:decomp_one}
\end{equation}
In the decomposition (\ref{eq:decomp_one}) the SVD basis $\{\tilde{\bar X}\}$ is truncated 
by disregarding the eigenvectors that correspond to eigenvalues smaller than a threshold: $\bar{D}_{\tilde{\bar X}}<\varepsilon_{T2}$.

Next, triples amplitudes in the mixed -- orbital and doubles-SVD -- basis are computed.
For this purpose we canonicalize the latter by diagonalizing the Fock matrix 
in the SVD basis $f_{\tilde{\bar X}}^{\tilde{\bar Y}}$:
\begin{equation}
f_{\tilde{\bar X}}^{\tilde{\bar Y}}=\bar{U}_{i\tilde{\bar X}}^{\dagger a}
(\epsilon_a - \epsilon_i)\bar{U}^{i\tilde{\bar Y}}_{a}.
\label{eq:fock_svd}
\end{equation}
In the basis of its eigenvectors $U_{\tilde{\bar X}}^{\bar{X}}$ the Fock matrix (\ref{eq:fock_svd})
is diagonal
\begin{equation}
f^{{\bar X}}_{{\bar Y}}=\epsilon_{\bar X}\delta^{{\bar X}}_{{\bar Y}}
\label{eq:fock_svd2}
\end{equation}
and the complete transformation to the ${\bar X}$-basis is given by
\begin{equation}
\bar{U}^{i{\bar X}}_{a}=\bar{U}^{i\tilde{\bar X}}_{a}\bar{U}_{\tilde{\bar X}}^{\bar{X}}.
\label{eq:U_bar}
\end{equation}
Having the transformation matrices to the (doubles) SVD basis at hand we can
obtain a partially decomposed CC3-type triples residual
\begin{align}
  R^i_{a\bar X\bar Y} = \left( X_{bX}^{jL} \bar{U}^{\dagger b}_{j \bar Y} \right) \hat{v}_a^{iL} 
  + V_{a \bar X}^d T^i_{d \bar Y} - V^i_{l \bar X} T^l_{a \bar Y} 
  + E^d_{l \bar X \bar Y} T_{ad}^{il}
  \label{eq:half_decomp_res}
\end{align}
with the intermediates
\begin{align}
  &X_{bX}^{jL} = T^j_{cX} \hat v_{b}^{cL} - T^l_{bX} \hat v_{l}^{jL}
  &T_{a \bar X}^i = \bar{U}^{\dagger b}_{j \bar X} T^{ij}_{ab} \\
  &E^d_{l \bar X \bar Y} = V_{a \bar X}^d \bar{U}_{l \bar Y}^{\dagger a} - V_{l \bar X}^{i} \bar{U}_{i \bar Y}^{\dagger d}
  &W_{\bar X}^L = \hat{v}_c^{kL} \bar{U}^{\dagger c}_{k \bar X} \\
  &V^d_{a \bar X} = \hat{v}_a^{dL} W^L_{\bar X}
  &V^j_{l \bar X} = \hat{v}^{jL}_l W^L_{\bar X}
\end{align}
at the $N^6$ cost at most. This scaling corresponds to the last term of
eq. (\ref{eq:half_decomp_res}), while all other terms scale even as $N^5$.
Since the SVD basis is canonicalized, the partially
transformed triples amplitudes are readily available:
\begin{eqnarray}
T^i_{a\bar X\bar Y} & =& \;- \frac{R^i_{a\bar X\bar Y} + R^i_{a\bar Y\bar X}}{(\epsilon_{\bar X} + \epsilon_{\bar Y} + \epsilon_{a} - \epsilon_{i})}\label{eq:HT_triples}
\end{eqnarray}
and with that -- the triples two-particle density matrix
\begin{eqnarray}
  D^{aj}_{ib} &=& T^{\dagger a \bar X\bar Y}_{i} T_{b \bar X\bar Y}^{j}\label{eq:Dijab}
\end{eqnarray}
again at the $N^6$ cost.

In the SVD basis the triples amplitudes implicitly include nonphysical
components with triply repeated occupied orbital
indices like $T_{abc}^{iii}$ which would cancel in case of a full SVD basis.
However, it is possible to explicitly isolate and discard their contributions to the density matrix:\\
\begin{eqnarray}
  \Delta D^{aj}_{ib} = (T^{\dagger a}_{i \bar X \bar Y} \mathcal{U}^{\bar X}_{i \bar X'} \mathcal{U}^{\bar Y}_{i\bar Y'}) T^{j \bar X' \bar Y'}_{b}
 \end{eqnarray}
with 
\begin{equation}
  \mathcal{U}^{\bar X }_{i\bar X'} = \sum_{a} \bar U^{\dagger a}_{i \bar X'} \bar U^{i \bar X}_{a}
\end{equation}
such that
\begin{eqnarray}
\widetilde{D}^{aj}_{ib} &=& D^{aj}_{ib} - \left(1 - \frac{1}{2}\delta^i_j \right) \mathcal{S}(ij,ab) \Delta D^{aj}_{ib}.
\end{eqnarray}

Finally, we obtain the actual SVD basis for the triples in DC-CCSDT by diagonalizing the triples two-electron 
density matrix $\widetilde{D}^{aj}_{ib}$
\begin{equation}
\widetilde{D}^{aj}_{ib} \approx U^{\dagger a}_{i \tilde{X}} D_{\tilde{X}}
U^{j\tilde{X}}_{b}.
\label{eq:decomp_two}
\end{equation}
Analogously to the doubles SVD, we disregard all the vectors whose 
eigenvalues are smaller than the chosen threshold:
$D_{\tilde{X}}<\varepsilon_{T3}$. Again, before the use in DC-CCSDT the
SVD basis is canonicalized by diagonalizing the Fock matrix,
\begin{equation}
f_{\tilde{X}}^{\tilde{Y}}=U_{i\tilde{X}}^{\dagger a}
(\epsilon_a - \epsilon_i)U^{i\tilde{Y}}_{a}.
\label{eq:fock_svd3}
\end{equation}
The eigenvectors $U_{\tilde{X}}^{X}$ of $f_{\tilde{X}}^{\tilde{Y}}$ are used in the final SVD transformation matrices,
\begin{equation}
U^{iX}_{a}=U^{i\tilde{X}}_{a}\bar{U}_{\tilde{X}}^{X}.
\label{eq:U}
\end{equation}
The matrices $U^{iX}_{a}$ of eq. (\ref{eq:U}) are the ones that are used in the
DC-CCSDT residual equations given in section \ref{sec:resid}.  

The additional SVD basis for the projection approximation, Eq.~(\ref{eq:proj}),
is contructed by half-transforming the contravariant doubles amplitudes, $\tilde T^{ij}_{ab}$, 
to the SVD basis, projecting them from full $\tilde T^{ij}_{ab}$,
\begin{align}
  \Delta \tilde T^{ij}_{ab} = \tilde T^{ij}_{ab} - U^{iX}_{a} U^{\dagger c}_{kX} \tilde T^{kj}_{cb},
\end{align}
computing SVD of $\Delta \tilde T^{ij}_{ab}$, and selecting the most important (left) vectors,
$\Delta U^{i\bar X}_a$, which are then added to the triples SVD basis. Finally,
we perform a symmetric orthogonalization of the space.

The eigenvalues $\epsilon_{X}$ of $f_{\tilde{X}}^{\tilde{Y}}$ are employed in the
update, which is carried out directly in the SVD basis:
\begin{equation}
\Delta T_{XYZ} = -\frac{R_{XYZ}}{\epsilon_X + \epsilon_Y + \epsilon_Z}.
\end{equation}
Due to the symmetry property $T_{XYZ} = T_{YXZ} = T_{XZY} = T_{ZYX} = T_{ZXY}
= T_{YZX}$ one can restrict solving the residual equations to
$X\leq Y \leq Z$. However, since our implementation is based on the implicit
tensor contraction library \texttt{TensorOperations.jl}\cite{TensorOperations.jl} 
we keep the full tensors.
To accelerate the convergence of DC-CCSDT 
we employ the direct inversion of the iterative subspace (DIIS) for singles, doubles and triples amplitude vectors.

\section{Test Calculations} \label{sec:half_decomp_res}

To test the performance of SVD-DC-CCSDT we benchmark it against DC-CCSDT and CCSDT(Q) (not density-fitted) 
for a set of 35 reaction energies from Ref. \onlinecite{huntington2012accurate}
(we exclude the LiH formation reaction as it contains only two-electron systems in frozen-core approximation).
Below we focus on the mean absolute (MEAN), the root-mean-square (RMS) and the maximum (MAX) deviations for the complete set. 
The detailed compilation of the results for each individual reaction can be found in the supplementary material.

All CCSD(T), DC-CCSDT, SVD-CCSD(T) and SVD-DC-CCSDT calculations employ density fitting and 
are performed using \texttt{ElemCo.jl}\cite{elemcojl}, a \texttt{Julia}-based quantum chemistry package. 
\texttt{ElemCo.jl} includes an interface to the \texttt{libcint}\cite{libcint} library for the electron-repulsion integrals. 
In the calculations we use the orbital basis sets from the Dunning basis set family.\cite{dunning1989gaussian} 
In the post-HF treatment the core is not correlated, and the density fitting approximation employs the respective auxiliary basis sets.\cite{weigend2002a} 
In the density fitted HF the def2-universal-jkfit auxiliary basis set\cite{weigend2008a} is chosen.


Firstly, we investigate the effect of the truncation of the SVD basis for
the CCSD doubles amplitudes, eq. (\ref{eq:decomp_one}), used to calculate the triples' density matrix for the subsequent decomposition of the triples. 
This truncation, governed by the threshold $\varepsilon_{T2}$, is an additional approximation to the SVD-DC-CCSDT method. 
Therefore, it is important to keep its influence negligible compared to that of the truncation of the SVD basis for the triples. 
In order to guarantee that by tightening of the triples SVD threshold $\varepsilon_{T3}$ the threshold $\varepsilon_{T2}$ would automatically be tightened too,
we link the two by a factor $w_{T2}$:
\begin{eqnarray}
\varepsilon_{T2}=w_{T2}\varepsilon_{T3}.\label{eq:tol_weight}
\end{eqnarray}

\begin{figure}[h]
  \centering
  \includegraphics[width=7cm]{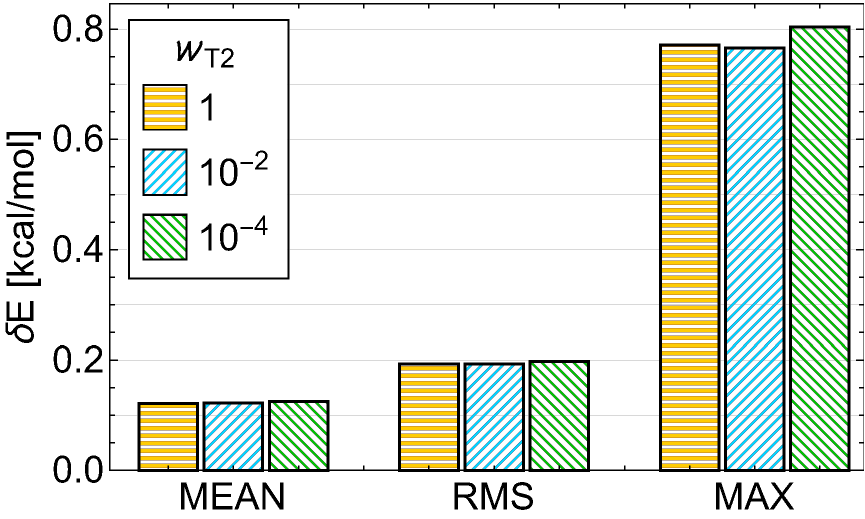} \hspace{0.7cm}
  \caption{    The mean, RMS and maximum deviations of the SVD-DC-CCSDT reaction energies as functions of the $w_{T2}$ weight factor for the $\varepsilon_{T2}$ tolerance (see eq. (\ref{eq:tol_weight})) with respect to the CCSDT(Q) reference.  
  The calculations employ the cc-pVDZ basis and fixed $\varepsilon_{T3} = 10^{-6}$.}
\label{fig:w2_dependence}
\end{figure}

Fig. \ref{fig:w2_dependence} presents the deviations of the SVD-DC-CCSDT energies from the CCSDT(Q) 
reference calculated with different values of $w_{T2}$ -- from 10$^{-4}$ to 1. 
Surprisingly, the accuracy of the method is insensitive to the choice of  $w_{T2}$. 
A likely reason of such a minute effect of the truncation of the SVD basis for the doubles is 
only a partial representation of the triples residuals (\ref{eq:half_decomp_res}) and the corresponding 
triples amplitudes  (\ref{eq:HT_triples}) in the truncated SVD basis. 
Therefore, the two-particle density matrix (\ref{eq:Dijab}) still contains information about the complete orbital space. 

\begin{figure*}[!ht]
  \begin{tabular}{cc}
    \includegraphics[width=7.8cm]{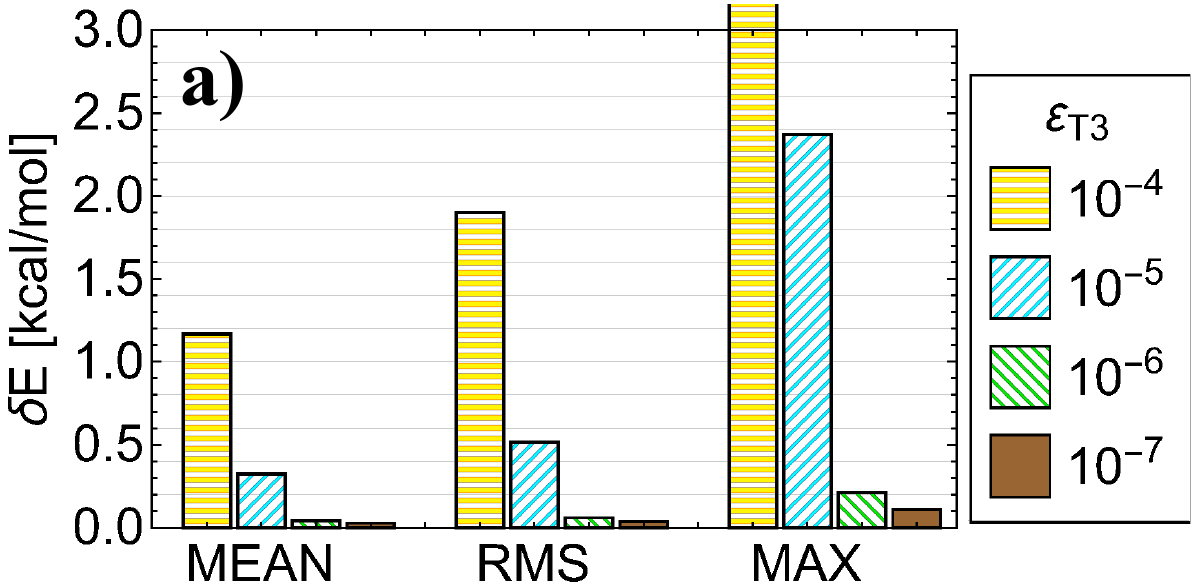} \hspace{0.6cm} & \includegraphics[width=7.8cm]{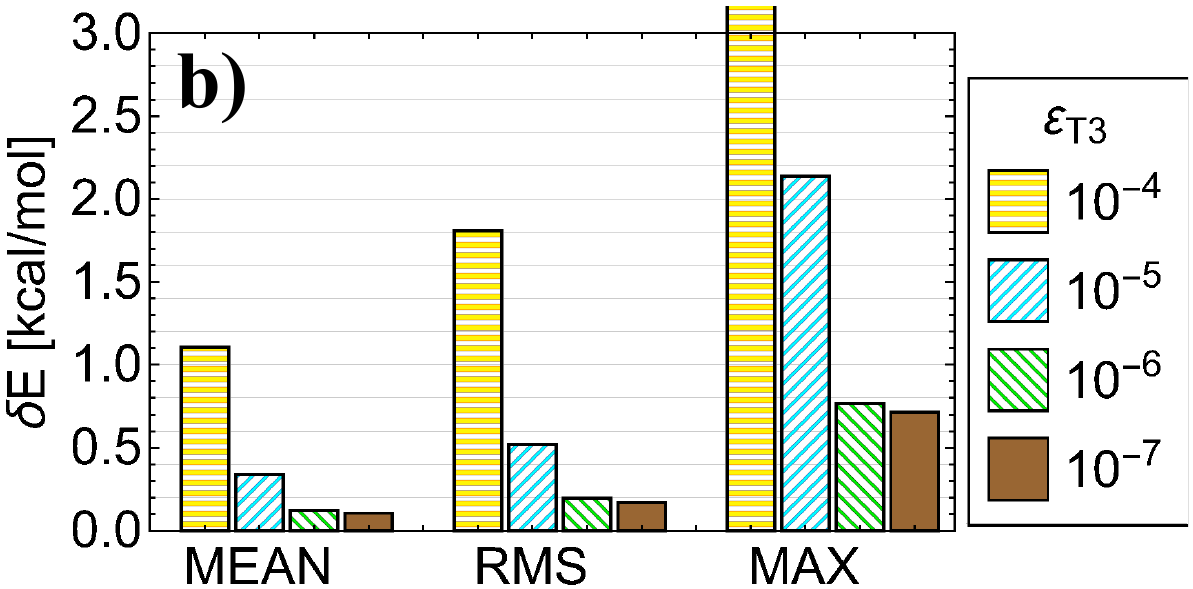} \hspace{0.6cm} \\
    \hspace{0.2cm}\includegraphics[width=8.7cm]{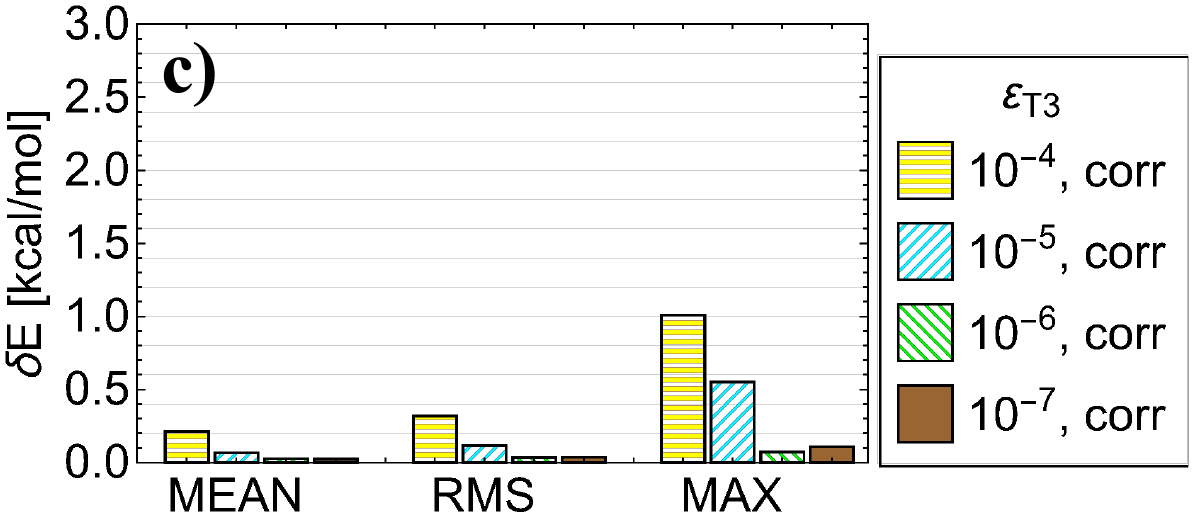} & \hspace{0.1cm} \includegraphics[width=8.6cm]{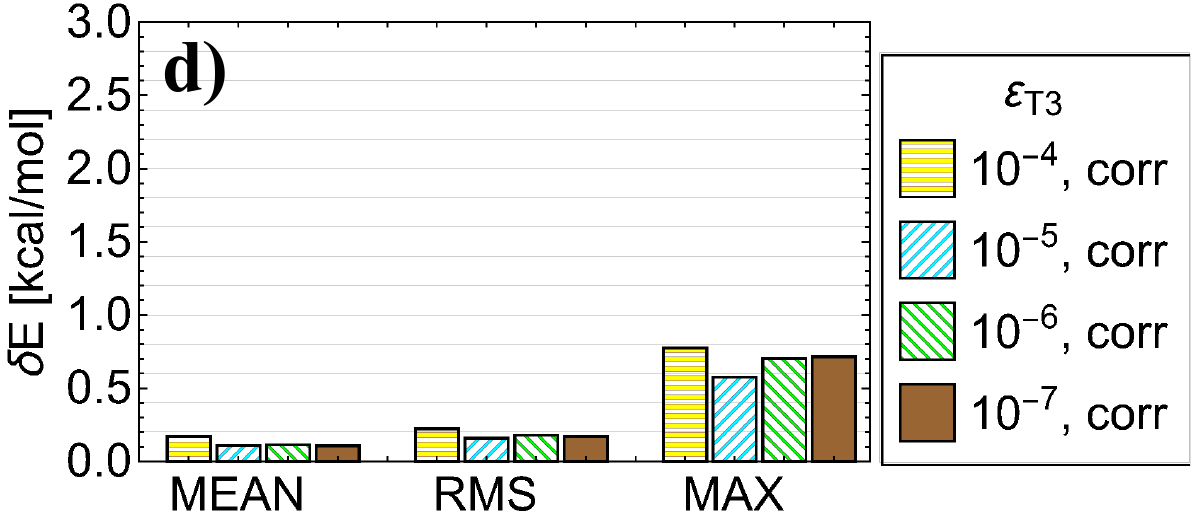}
  \end{tabular}
  \caption{
    The mean, RMS and maximum deviations of the SVD-DC-CCSDT reaction energies as functions 
    of the $\varepsilon_{T3}$ tolerance with respect to the DC-CCSDT (panels a and c) and 
    CCSDT(Q) (panels b and d) references. 
    The calculations employ the cc-pVDZ basis and fixed $w_{T2} = 10^{-2}$. 
    In panels c and d, the SVD-error was corrected using the hybrid approach of eqs. (\ref{eq:corrected}) 
    and (\ref{eq:corrected2}).
      }
\label{fig:eps3_dependence}
\end{figure*}

Next, we focus on the main approximation of the method -- the truncation of the triples SVD space. 
In Fig. \ref{fig:eps3_dependence}a it is demonstrated that the SVD-DC-CCSDT reaction energies for 
the test set calculated with the cc-pVDZ basis, 
progressively approach the untruncated DC-CCSDT ones with tightening of the threshold $\varepsilon_{T3}$. 
The appreciable accuracy (the mean and RMS deviations $<0.1$ kcal/mol, maximum deviation $~0.2$ kcal/mol), however,
can only be achieved with $\varepsilon_{T3}$ not higher than 10$^{-6}$. 
A comparison to the higher-level CCSDT(Q) results, shown in Fig. \ref{fig:eps3_dependence}b reveals 
that at this value of $\varepsilon_{T3}$ the truncation error becomes smaller than the DC-CCSDT method error.

In order to accelerate convergence with the SVD truncation threshold we tested a hybrid approach 
with the SVD-error correction at the CCSD(T) level:
\begin{equation}
  E_{\textrm{DC-CCSDT}} \approx E_{\textrm{SVD-DC-CCSDT}}  + \delta_{\rm SVD}\label{eq:corrected}
\end{equation}
with
\begin{equation}
  \delta_{\rm SVD}=E_{\textrm{CCSD(T)}} - E_{\textrm{SVD-CCSD(T)}}.\label{eq:corrected2}
\end{equation}
As is seen in Fig. \ref{fig:eps3_dependence}c, the hybrid approach (\ref{eq:corrected}) indeed converges noticeably faster. 
A sufficient accuracy is achieved already with $\varepsilon_{T3} = 10^{-5}$. 
Interestingly, comparison to CCSDT(Q) in Fig. \ref{fig:eps3_dependence}d shows that the SVD-error 
in the hybrid scheme is no longer  dominant even for $\varepsilon_{T3} = 10^{-4}$.
The CCSD(T) energy in Eq.~(\ref{eq:corrected2}) can be replaced by SVD-CCSD(T) energies calculated 
using one order of magnitude tighter threshold. 
In our test calculations the accuracy of this correction scheme is very close to using 
the complete CCSD(T) energies (see Supplementary Information).

\begin{figure}[!h]
  \centering
  \includegraphics[width=8cm]{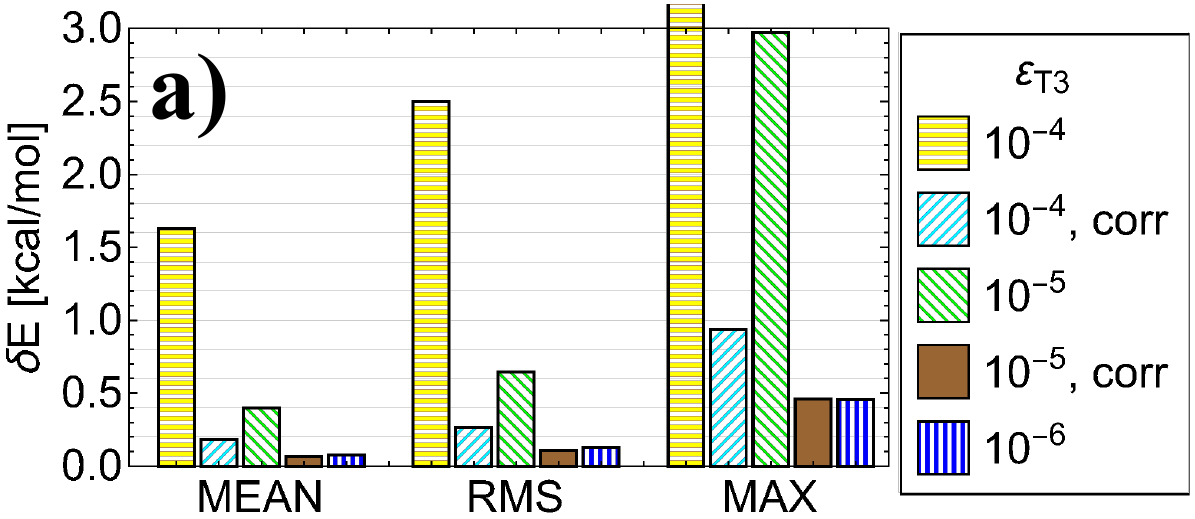}
  \\\includegraphics[width=8cm]{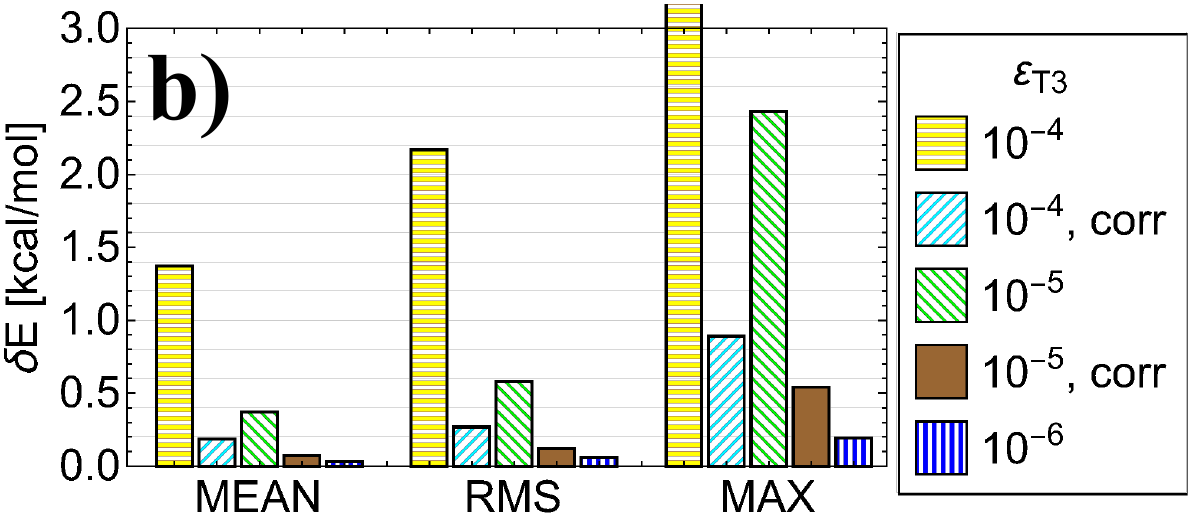}
  \caption{  The mean, RMS and maximum deviations of the SVD-DC-CCSDT reaction energies as 
  functions of the $\varepsilon_{T3}$ tolerance, calculated with and without the hybrid approach 
  (see eqs. (\ref{eq:corrected}) and (\ref{eq:corrected2})) and two different basis sets: 
  aug-cc-pVDZ (panel a) and cc-pVTZ (panel b). 
  The reference results were obtained with the hybrid approach and  $\varepsilon_{T3}=10^{-6}$. 
  In all the calculations the weight $w_{T2}$ of $10^{-2}$ was chosen.}
 \label{fig:basis_set_quality}
\end{figure}

A qualitatively similar picture is observed for larger basis sets -- aug-cc-pVDZ and cc-pVTZ -- 
shown in Fig. \ref{fig:basis_set_quality}. 
Due to the computational cost, for these basis sets not only the CCSDT(Q) reference is unfeasible 
but also the untruncated DC-CCSDT one. 
So here we use the SVD-DC-CCSDT results with  $\varepsilon_{T3} = 10^{-6}$ obtained with the 
hybrid scheme (\ref{eq:corrected}) as the reference. 
Again the pure SVD-DC-CCSDT is reliable, provided the threshold $\varepsilon_{T3}$ is tightened to  10$^{-6}$. 
With the hybrid approach, however, a comparable accuracy is reached already with $\varepsilon_{T3} = 10^{-5}$. 
Moreover, even with a very crude tolerance of $\varepsilon_{T3} = 10^{-4}$ the RMS and mean 
deviations of the hybrid scheme are about 0.2 kcal/mol and the maximum deviation for the 
reaction set is smaller than 1 kcal/mol. 

A possibility to use looser  $\varepsilon_{T3}$ thresholds, provided by the hybrid approach, 
is essential, as it has a strong impact on the computational cost.
In table \ref{tab:timings} we list the sizes of the SVD space and timings of the SVD-DC-CCSDT calculations  
(within the hybrid scheme) on a benzene molecule with different basis sets. 
Tightening of  $\varepsilon_{T3}$ by an order of magnitude results in an increase of the time 
for SVD-DC-CCSDT iterations also by an order of magnitude on average. 
Furthermore, for the threshold values of 10$^{-4}$ or 10$^{-5}$ the cost of the SVD-DC-CCSDT 
calculations is lower or comparable to that of the untruncated (T) needed for the hybrid scheme (\ref{eq:corrected}). Due to the higher scaling of (T) compared to SVD-DC-CCSDT,  the former is expected to become the actual bottleneck of hybrid scheme calculations on larger systems, which can be circumvented by the use of SVD-(T) with tighter $\varepsilon_{T3}$ instead, as discussed above. It is also worth noting that the size of the SVD basis after truncation within a given threshold does grow 
with the orbital basis but very slowly, much slower than the number of virtual orbitals.

\begin{table}[h]
  \caption{The number of the SVD basis functions ($n_{\textrm{SVD}}$) and respective timings 
  for SVD-DC-CCSDT calculations on the benzene molecule from the test set with different 
  $\varepsilon_{T3}$ thresholds and cc-pVDZ, aug-cc-pVDZ and cc-pVTZ basis sets.
  $t_{\textrm{DEC}}$ is the time taken for the construction of the triples SVD basis.
  All timings are in minutes.
  In bold, we highlight the results for the recommended $\varepsilon_{T3}=10^{-5}$ threshold.
  The number of virtual orbitals for each basis set is denoted by $n_{\textrm{virt}}$. 
  The weight $w_{T2}=10^{-2}$ was chosen.
  All calculations were performed on 12-core Intel Xeon Gold 6128 nodes with the rate of 3.40GHz.}
  \begin{ruledtabular}
  \begin{tabular}{lccccc}
   & $\varepsilon_{T3}$ & $n_{\textrm{SVD}}$ & $t_{\textrm{DEC}}$ & $t_{\textrm{SVD-(T)}}$ & $t_{\textrm{SVD-DC-CCSDT}}$ \\
  \hline
  \bf{cc-pVDZ} & $10^{-4}$ & 33 & 0.19 &  0.01 & 0.33 \\
  ($n_{\textrm{virt}} = 93$)& $\mathbf{10^{-5}}$ & \textbf{147} & \textbf{0.36} & \textbf{0.07} & \textbf{1.33} \\
  ($t_{\textrm{CCSD}} = 0.19$)& $10^{-6}$ & 440 & 0.69 & 0.38 & 19.90 \\
  ($t_{\textrm{(T)}} = 0.44$)& $10^{-7}$ & 950 & 1.23 & 2.00 & 430.09 \\
  \hline
  \bf{aug-cc-pVDZ}& $10^{-4}$ & 36 & 0.62 & 0.04 & 1.47 \\
  ($n_{\textrm{virt}} = 171$)& $\mathbf{10^{-5}}$ & \textbf{161} & \textbf{1.29} & \textbf{0.18} & \textbf{4.14} \\
  ($t_{\textrm{CCSD}} = 0.86$)& $10^{-6}$ & 494 & 2.65 & 1.03 & 44.33 \\
  ($t_{\textrm{(T)}} = 3.30$)& $10^{-7}$ & 1126 & 5.37 & 5.31 & 933.60 \\
  \hline
  \bf{cc-pVTZ}& $10^{-4}$ & 36 & 1.79 & 0.08 & 3.88 \\
  ($n_{\textrm{virt}} = 243$)& $\mathbf{10^{-5}}$ & \textbf{187} & \textbf{3.60} & \textbf{0.40} & \textbf{10.00} \\
  ($t_{\textrm{CCSD}} = 2.45$)& $10^{-6}$ & 592 & 7.90 & 2.33 & 91.57 \\
  ($t_{\textrm{(T)}} = 12.47$) & $10^{-7}$ & 1450 & 17.21 & 14.01 & 3431.22 \\
  \end{tabular}
  \end{ruledtabular}
\label{tab:timings}
\end{table}

Finally, we investigate how the size of the truncated SVD basis and the truncation error scale with the system size. 
The theoretical $N^6$ scaling that formally follows from the equations (\ref{eq:resid_trip})-(\ref{eq:int4}) 
and (\ref{eq:half_decomp_res})-(\ref{eq:Dijab}) will be reproduced in practice only if the number 
of the SVD basis functions after truncation grows linearly with systems size. 
Furthermore, for the size-extensivity of the method it is vital that the SVD-truncation error 
also scales linearly with the systems size. 
In order to test these two aspects we considered a set of formal linear alkanes from $(\textrm{CH}_2)_2 \textrm{H}_2$ 
to $(\textrm{CH}_2)_5 \textrm{H}_2$ with the following parameters: 
$r_{\textrm{C-C}} = 1.54$ {\AA}, $r_{\textrm{C-H}} = 1.1$ {\AA} and $\alpha_{\textrm{C-C-C}} = \alpha_{\textrm{H-C-H}} = 109.5^{\circ}$.

\begin{figure}
    \hspace{0.2cm} \includegraphics[width=6cm]{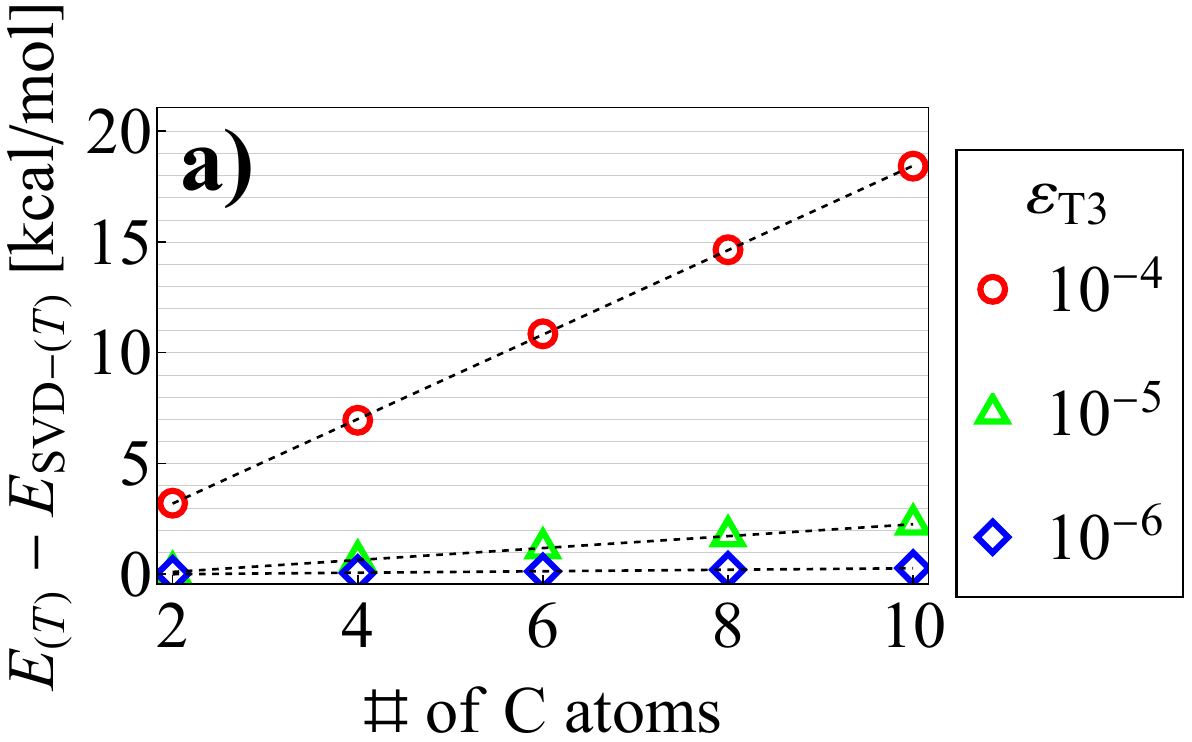} 
    \\ \includegraphics[width=6.2cm]{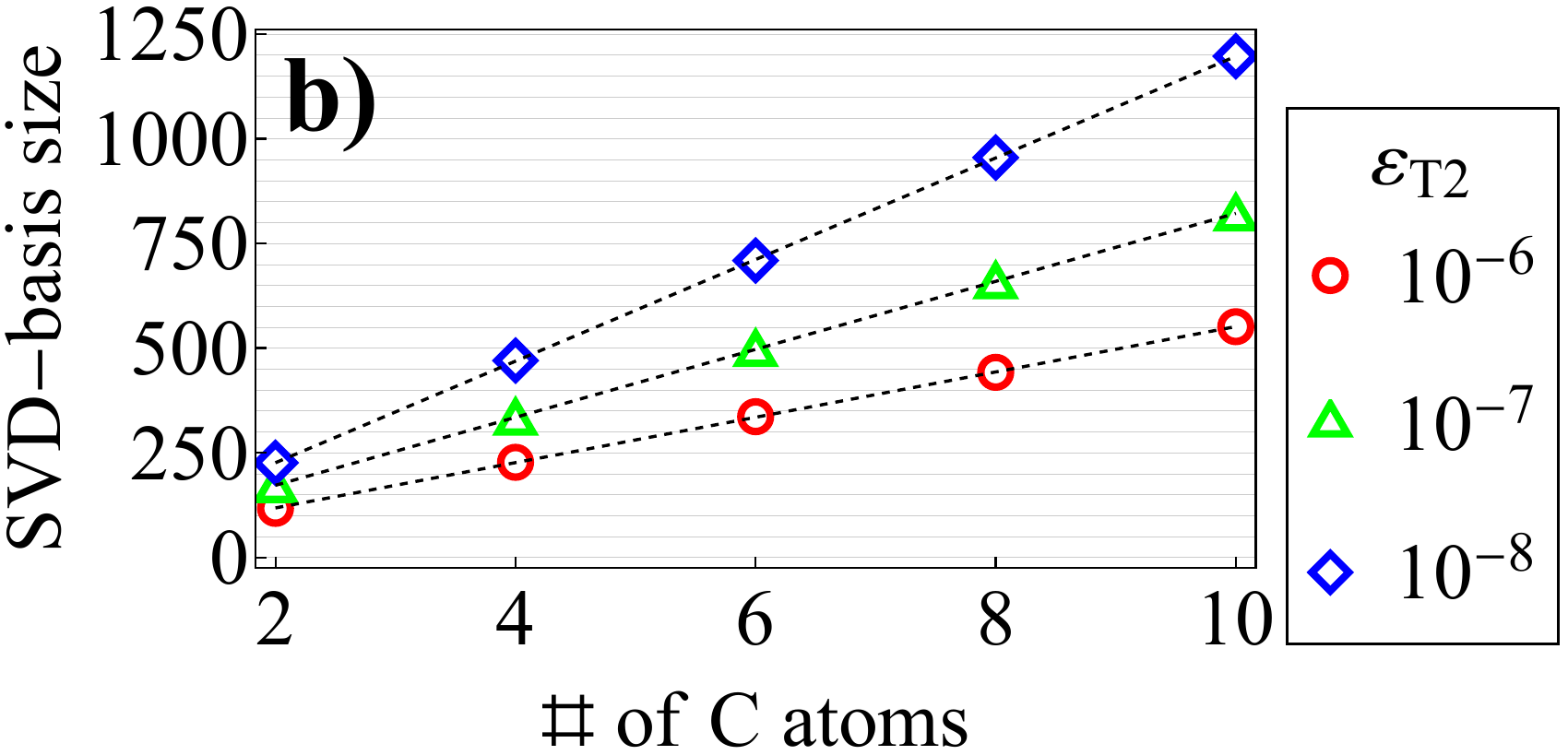}
    \\ \hspace{0.1cm} \includegraphics[width=6.0cm, height=3.cm]{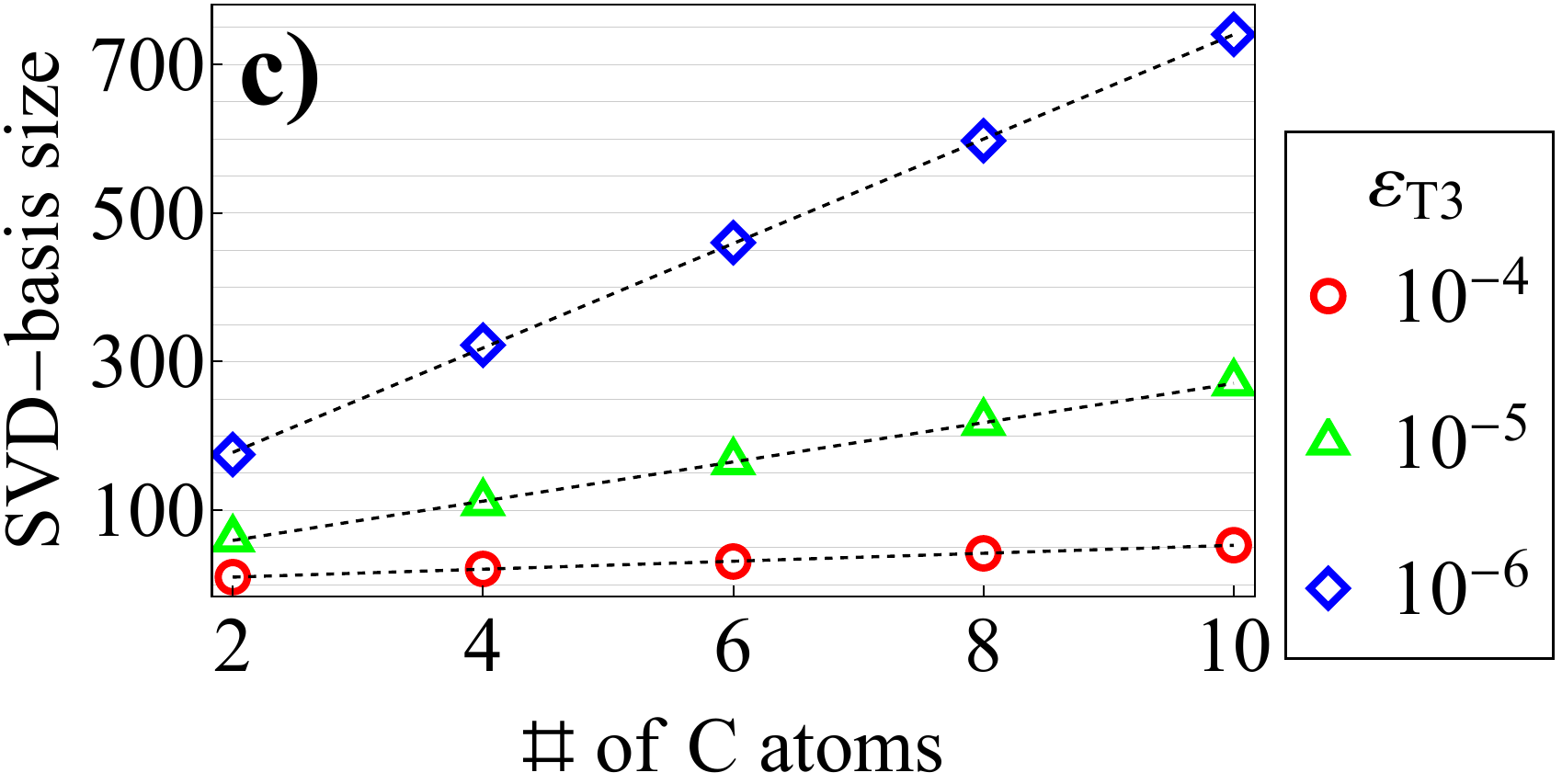}
    \caption{The SVD-error of the SVD-(T) energy (panel a) and the SVD basis sizes as functions 
    of the length of linear  alkane chains
    from $(\textrm{CH}_2)_2 \textrm{H}_2$ to $(\textrm{CH}_2)_5 \textrm{H}_2$, 
    calculated with different thresholds $\varepsilon_{T3}$. 
    Panel b reports the sizes of the SVD-space for the doubles decomposition 
    (see eqs.  (\ref{eq:decomp_one})-(\ref{eq:Dijab})), 
    the panel c -- the SVD space for the triples decomposition (see eqs. (\ref{eq:decomp_two})-(\ref{eq:U}))}.
\label{fig:size_extensivity}
\end{figure}

Firstly, we focus on the size-extensivity of the SVD truncation error. 
Since the DC-CCSDT calculations for larger alkanes are too expensive, we evaluate this error at the SVD-(T) level. 
As is seen from Fig. \ref{fig:size_extensivity}a the error is clearly linear with the system size for all the thresholds used, 
confirming the size-extensivity of the truncation strategy based on the cutoff tolerance. 
As concerns the error itself, with $\varepsilon_{T3}=10^{-4}$ it is clearly unacceptably large. 
With $\varepsilon_{T3}=10^{-6}$ the errors are very small: for the largest alkane considered, 
the error in the total (T) energy is below 0.3 kcal/mol. 
The choice $\varepsilon_{T3}=10^{-5}$ leads to small errors, but not necessarily negligible for larger systems. 
However, as established above, the hybrid scheme reduces this error to appreciable levels. 
Finally, the linear growth of the truncated SVD-spaces is also reproduced, 
as is evident from Figs. \ref{fig:size_extensivity}b and \ref{fig:size_extensivity}c.


\section{Conclusions}

In this work we presented an SVD-DC-CCSDT method, which is a very accurate but much less expensive 
approximation to DC-CCSDT.
In SVD-DC-CCSDT the triples amplitudes as well as the triples residuals are represented in 
the truncated basis of singular vectors. 
At no point in the method the full triples amplitudes in the molecular orbital basis need 
to be processed.  
The factorization of the intermediates in the triples residual equations achieved by 
the decomposition of the triples amplitudes and the density fitting reduces the scaling 
from ${\cal O}(N^8)$ to ${\cal O}(N^6)$.
The DC approximation not only improves the accuracy of the CCSDT method,
but also removes the most expensive terms from the SVD-CCSDT method.

The determination of the SVD transformation matrices is evaluated via the CC3-like triples 
contribution to the 2-particle density matrix. 
This procedure also scales as ${\cal O}(N^6)$ due to a partial transformation of the involved 
intermediates in an auxiliary SVD basis for the doubles amplitudes.   

The truncation of both SVD spaces are governed by the tolerances $\varepsilon_{T3}$ for the triples 
and $\varepsilon_{T2}$ in the auxiliary SVD for the doubles. 
These thresholds are compared to the eigenvalues of the respective two-electron density matrices. 
The choice of $\varepsilon_{T2}$ has a very little effect on the results, at least if it is 
not larger than $\varepsilon_{T3}$. 
At the same time, the SVD-error depends on the choice of $\varepsilon_{T3}$ itself quite strongly. 
By benchmarking reaction energies for a test set of 42 small molecules, it was determined 
that with the $\varepsilon_{T3}=10^{-6}$ the error of the SVD approximation is smaller than 
the error of the DC-CCSDT method itself compared to CCSDT(Q). 
A correction of the SVD-error at the SVD-(T) level allows one to loosen this tolerance to $10^{-5}$. 
Regardless of the choice of $\varepsilon_{T3}$, the SVD-error in the energy and the size of 
the SVD bases scale linearly with the system size.


\begin{acknowledgments}
C.R. thanks Studienstiftung des deutschen Volkes for a masters scholarship.
\end{acknowledgments}

\bibliography{Paper}

\end{document}